\documentclass[fleqn, usenatbib]{mnras}

\usepackage{newtxtext,newtxmath}

\usepackage[T1]{fontenc}
\usepackage{ae,aecompl}
\usepackage{graphicx}
\usepackage{blkarray}
\usepackage{verbatim}

\usepackage{graphicx}	
\usepackage{amsmath}	
\usepackage{amssymb}	
\usepackage{amsfonts}
\usepackage{color,soul}
\makeatletter
\newcommand*\bigcdot{\mathpalette\bigcdot@{.5}}
\newcommand*\bigcdot@[2]{\mathbin{\vcenter{\hbox{\scalebox{#2}{$\m@th#1\bullet$}}}}}
\makeatother

\title[RGZ: deep learning for radio source classification]{Radio Galaxy Zoo: \cla\ --- a deep learning classifier for radio morphologies}

\author[C Wu et al.]{Chen Wu$^{1}$\thanks{E-mail: chen.wu@icrar.org (ICRAR)},
O.\ Ivy Wong$^{1}$\thanks{E-mail: ivy.wong@uwa.edu.au}, Lawrence Rudnick$^{2}$, 
 Stanislav S. Shabala$^{3}$
\newauthor Matthew J. Alger$^{4,5}$, Julie K. Banfield$^{4,6}$, Cheng Soon Ong$^{5}$, Sarah V. White$^{7}$,  
\newauthor Avery F. Garon$^{2}$, Ray P. Norris$^{8,9}$, Heinz Andernach$^{10}$, Jean Tate$^{11}$
\newauthor Vesna Lukic$^{12}$, Hongming Tang$^{13}$, Kevin Schawinski$^{14}$, Foivos I. Diakogiannis$^{15,1}$
\\
$^{1}$International Centre for Radio Astronomy Research (ICRAR), The University of Western Australia, 35 Stirling Hwy, Crawley, WA 6009, Australia\\
$^{2}$School of Physics and Astronomy, University of Minnesota, Minneapolis, MN 55455, USA\\
$^{3}$School of Natural Sciences, University of Tasmania, Private Bag 37, Hobart, Tasmania 7001, Australia\\
$^{4}$Research School of Astronomy and Astrophysics, The Australian National University, Canberra, ACT 2611, Australia\\
$^{5}$Data61, CSIRO, Canberra, ACT 2601, Australia\\
$^{6}$ARC Centre of Excellence for All-Sky Astrophysics (CAASTRO), Building A28, School of Physics, The University of Sydney, NSW 2006, Australia\\
$^{7}$International Centre for Radio Astronomy Research (ICRAR), Curtin University, Bentley, WA 6102, Australia\\
$^{8}$Western Sydney University, Locked Bag 1797, Penrith South, NSW 1797, Australia\\
$^{9}$CSIRO Astronomy and Space Science, Australia Telescope National Facility, PO Box 76, Epping, NSW 1710, Australia\\
$^{10}$Depto. de Astronom\'{i}a, DCNE, Universidad de Guanajuato, Apdo. Postal 144, Guanajuato, CP 36000, Gto., Mexico\\
$^{11}$Zooniverse Citizen Scientist, c/o Oxford Astrophysics, Denys Wilkinson Building, Keble Road, Oxford OX1 3RH, UK\\
$^{12}$Hamburger Sternwarte, University of Hamburg, Gojenbergsweg 112, D-21029 Hamburg, Germany\\
$^{13}$School of Physics and Astronomy, University of Manchester, Oxford Road, Manchester, M13 9PL, UK\\
$^{14}$Institute for Particle Physics and Astrophysics, ETH Z\"urich, Wolfgang-Pauli-Str. 27, CH-8093 Z\"urich, Switzerland\\
$^{15}$Data61, CSIRO, Floreat, WA 6014, Australia\\
}

\date{Accepted XXX. Received YYY; in original form ZZZ}

\pubyear{2018}

\newcommand{\cla}{{\textsc{ClaRAN}}}

\begin{document}

\label{firstpage}
\pagerange{\pageref{firstpage}--\pageref{lastpage}}
\maketitle

\begin{abstract}
The upcoming next-generation large area radio continuum surveys can expect tens of millions of radio sources, rendering the traditional method for radio morphology classification through visual inspection unfeasible. We present {\cla} --- \textbf{Cla}ssifying \textbf{R}adio sources \textbf{A}utomatically with \textbf{N}eural networks --- a proof-of-concept radio source morphology classifier based upon the Faster Region-based Convolutional Neutral Networks (Faster R-CNN) method.  Specifically, we train and test \cla\ on the FIRST and WISE images from the Radio Galaxy Zoo Data Release 1 catalogue. {\cla} provides end users with automated identification of radio source morphology classifications from a simple input of a radio image and a counterpart infrared image of the same region. {\cla} is the first open-source, end-to-end radio source morphology classifier that is capable of locating and associating discrete and extended components of radio sources in a fast (\(< 200\) milliseconds per image) and accurate ($\ge$ 90\%) fashion. 
 Future work will improve {\cla}'s relatively lower success rates in dealing with multi-source fields and will enable {\cla} to identify sources on much larger fields without loss in classification accuracy.

\end{abstract}




\section{Introduction}
Understanding the growth and evolution of Active Galactic Nuclei (AGN) is a fundamental area of research in the field of galaxy evolution as the pre-Square Kilometre Array (pre-SKA) experiments are now beginning their surveys. 
Radio AGN can be classed as `jetted' or `non-jetted'~\citep{padovani17}. 
On larger angular scales, radio jets can extend to great distances away from their host galaxies depending on their intrinsic mechanical energy and the environment into which they are launched. Over time, a bipolar jet may fade into two distinct radio lobes that are no longer connected to the host galaxy where it originated. Therefore, while approximately 90 percent of radio sources are compact in structure, 
 the remaining radio galaxy morphologies are extended with multiple radio source components and a rich set of structures.


Until now the cross-identification of  associated radio source components as well as the originating host galaxies are made via visual inspection. Currently, the most efficient form of visual identification is via citizen science projects such as Radio Galaxy Zoo \citep[RGZ; ][]{banfield15}. RGZ is based on large-area radio surveys and the efficacy of this project is demonstrated by the science results and recent discoveries  of extreme classes of radio source morphologies \citep{banfield16,kapinska17,contigiani17}.  
  
On the other hand, it is clear that we have reached even the limitations of citizen science since the number of complex, extended sources expected from the next-generation radio surveys such as the Evolutionary Map of the Universe \citep[EMU; ][]{norris11} will be far too great for a standalone citizen science project to be an efficient method. Therefore, automated methods of classification are necessary. Simple automated methods based upon source position matching can be effective for a significant fraction of radio sources \citep[e.g.\ ][]{kimball08}. However, complex extended radio sources with multiple discrete components and morphology will require more sophisticated methods. Therefore, Deep learning methods provide one such avenue for the specific task of radio source identification and classification.  Recently, \citet{wright17} demonstrated that a combination of citizen science and deep learning methods will maximize the science output of a dataset and outperform the capabilities of each method individually.
 
The main purpose of this paper is to present a proof-of-concept, publicly-available\footnote{\texttt{https://github.com/chenwuperth/rgz\_rcnn/}}, deep learning-based method known as Classifying Radio sources Automatically using Neural networks (\cla). {\cla} takes as input a pair of World Coordinate System-aligned radio and infrared images. It finds all radio sources and classifies them into one of the six morphology classes based on RGZ. The six classes of morphologies are not defined in the traditional manner of Fanaroff-Riley (FR) classes --- FR-I and FR-II \citep{fanaroff74,owen94} --- but in terms of source associations and identifications that are produced by RGZ's Data Release 1 (Wong et al., in preparation) represented as the number of components and peaks.
Therefore a single radio galaxy or radio source can be composed of one or more components and/or peaks. 
This paper builds upon RGZ's earlier exploration in combining the results from RGZ with advanced machine learning algorithms such as \citet{lukic18} and \citet{alger18}. 

We briefly introduce advanced machine learning (also known as deep learning) methods in Section \ref{sec:deep_learning}. The RGZ citizen science project and data pre-processing for feature fusion is described in Section \ref{sec:use_rgz_datasets}.  In the spirit of open source reproducibility, Section \ref{sec:data_pipeline} provides a complete technical description of \cla.  
Section \ref{sec:evaluation} details the error analysis and metrics-based evaluation commonly used in the field of machine learning. Section \ref{sec:empirical_eval} describes an example of the simplest automated application of \cla\ from the perspective of an astronomer and its reliability verification analysis. This ensures the accuracy of the classifications and provides additional information on the presence of multiple radio sources within the same image.  
Implications of our work and future research are briefly discussed in Section \ref{sec:implications} and we provide a summary of our results in Section \ref{sec:conclusions}.



\section{Deep learning methods}
\label{sec:deep_learning}
Deep learning methods~\citep{lecun2015deep}, particularly Convolutional Neural Networks (CNNs)~\citep{krizhevsky2012imagenet}, have recently achieved recognition capabilities that are comparable to or even better than humans in several visual recognition tasks, such as understanding traffic signs~\citep{ciregan2012multi}, identifying faces~\citep{taigman2014deepface}, and classifying general images~\citep{he2016deep}. 
CNNs have recently been explored to address a number of astrophysical problems such as: 1) effective identification of  exoplanet candidatess ~\citep{Shallue2018Kepler,pearson18}; 2)  the identification of gravitational lenses~\citep{Schaefer2018Lensing} and the estimation of strong gravitational lensing parameters~\citep{Hezaveh2017Lensing}; 3) automatic visual detection of galaxy structures such as galactic bars and mergers~\citep{abraham18,ackermann18}; 4) the determination of physical stellar parameters from optical stellar spectra \citep{fabbro18}; and 5) the identification of transients in real-time via image differencing \citep{sedaghat18}.

Despite many successful applications of CNNs, automated deep learning methods for localizing and classifying multi-component, multi-peak radio sources are still in their infancy. This has motivated our work in this paper. The winning solution~\citep{dieleman2015rotation} of the \textit{Galaxy Challenge}\footnote{\texttt{https://kaggle.com/c/galaxy-zoo-the-galaxy-challenge}} did utilize CNNs for accurate (\(>90\%\)) galaxy morphology classification. However, our work solves a very different problem from the \textit{Galaxy Challenge}: 
we need to determine the number of radio sources in a given field of view (FoV) or \textit{subject} (as is referred to within the RGZ project), each of which may contain multiple discrete source components. Such a determination is estimated from the combination of a radio continuum image and an infrared map in the same position. Moreover, we need to localize each detected radio source with a bounding box, and finally to predict the morphology class for each detected source with some probability. Our problem is also different from radio continuum source finders, which typically involve identifying individual source components that are above a certain signal-to-noise threshold~\citep{hancock12}. We need to group these components into one or more radio sources, and provide the morphology classification for each radio source.

The CNN method developed in \citet{Aniyan17} accurately classifies a FIRST radio source into FR and bent-tailed (BT) morphology classes. Although {\cla} is closely related to \citet{Aniyan17}, our research problem and method differ from \citet{Aniyan17}. {\cla} performs two tasks --- \textit{source identification} in a given field and \textit{morphology classification} for each identified source. These two tasks address very different issues, 
and {\cla} is trained to solve both tasks simultaneously in a single, end-to-end training pipeline. During testing, {\cla} finds both compact and extended radio sources in all possible locations on an image, and classifies each one of them into some morphology. In contrast, the \citet{Aniyan17} CNN classifier is trained to perform morphology classification \textit{only}. As such the input image is cut out from the main image during pre-processing, and is \textit{centered} at a known, given source. Moreover, while both {\cla} and \citet{Aniyan17} use radio images, {\cla} can also use infrared signals to significantly improve classification performance as shown in Section \ref{sec:evaluation}. The ability to integrate multi-wavelength datasets for automated source identification and morphology classification is unique to {\cla}.

\subsection{{\cla} Overview}
In this work, we use Faster R-CNN model~\citep{ren2017faster} as the basis to develop {\cla} for identifying multi-component/peak radio sources from DR1. This is because Faster R-CNN 
is intuitive to understand, flexible to augment, and most importantly, offers optimal trade-offs between robust accuracy and execution latency~\citep{huang2017speed}. As a result, {\cla} includes an end-to-end data pipeline that enables fast identification and classification of radio sources with a mean Average Precision\footnote{It should be noted that \textit{precision} here differs from the definition ~\citep{bevington03} in physical sciences} (mAP, which is formally defined in Section \ref{sec:metrics_eval}) of 83.6\% and an empirical accuracy above 90\%. In particular, we make several contributions to deep learning-based methods for RGZ:
 \begin{itemize}
  \item We develop and evaluate several 
  methods to combine radio emission and near-infrared maps for source identification. This paves the way for future work on optimal (e.g. adaptive, learning-based) integration of multi-wavelength datasets for automated source-matching and identification.
  \item We tailor and fine-tune the Faster R-CNN~\citep{ren2017faster} --- a state-of-the-art object detection deep learning model --- for effective radio source detection. To the best of our knowledge, latest research in object detection and computer vision has not yet been explored and utilized for radio source identification. 
  \item We augment the Faster R-CNN model by replacing its Region-of-Interest (RoI) cropping layer (\texttt{RoI pooling}) with differentiable affine transformations (\texttt{ST pooling}) based on the Spatial Transformer Network~\citep{jaderberg2015spatial}. Compared to the original Faster R-CNN model, training {\cla} becomes truly end-to-end --- all training errors are accounted for by the learning model within a single data pipeline. 
  \item We develop a transfer learning~\citep{yosinski2014transferable, ackermann18} strategy --- loading weights pre-trained on the ImageNet~\citep{deng2009imagenet} dataset and selectively controlling low-level convolutional kernels --- to significantly accelerate the training error convergence.
  \item We demonstrate that {\cla} can distinguish between six distinct classes of radio source morphologies using both machine learning metrics and empirical accuracy evaluation performed by radio astronomers. 
  \item We evaluate {\cla}'s scalability by showing its ability to identify radio sources with plausible classifications when the angular size in each direction of its input field is five times greater than what is available in the training set.
  
  Taken together, our study provides an excellent starting platform for developing future machine learning-based methods for wide-area radio continuum surveys.
\end{itemize}

\section{Using Radio Galaxy Zoo classifications}
\label{sec:use_rgz_datasets}
The citizen science project RGZ obtains visual identification of radio sources from over 12,000 volunteers, who have collectively completed over two million classifications to date.  Upon completion, RGZ will result in a catalogue of associated radio components and cross-matched host galaxies for over 170 thousand radio sources from the Faint Images of the Radio Sky at Twenty-centimeters \citep[FIRST; ][]{becker95} survey and over 2000 sources from the Australia Telescope Large Area  Survey \citep[ATLAS; ][]{norris2006deep}.  Currently, the cross-identification of extended radio sources and sources with disconnected radio lobes is through the visual inspection of radio sky maps with 
near-infrared maps. Therefore, the method of crowd-sourcing is used in RGZ to create one of the largest catalogues of extended radio galaxies with associated source components and host galaxy identifications. 
\begin{figure}
	\includegraphics[width=\columnwidth]{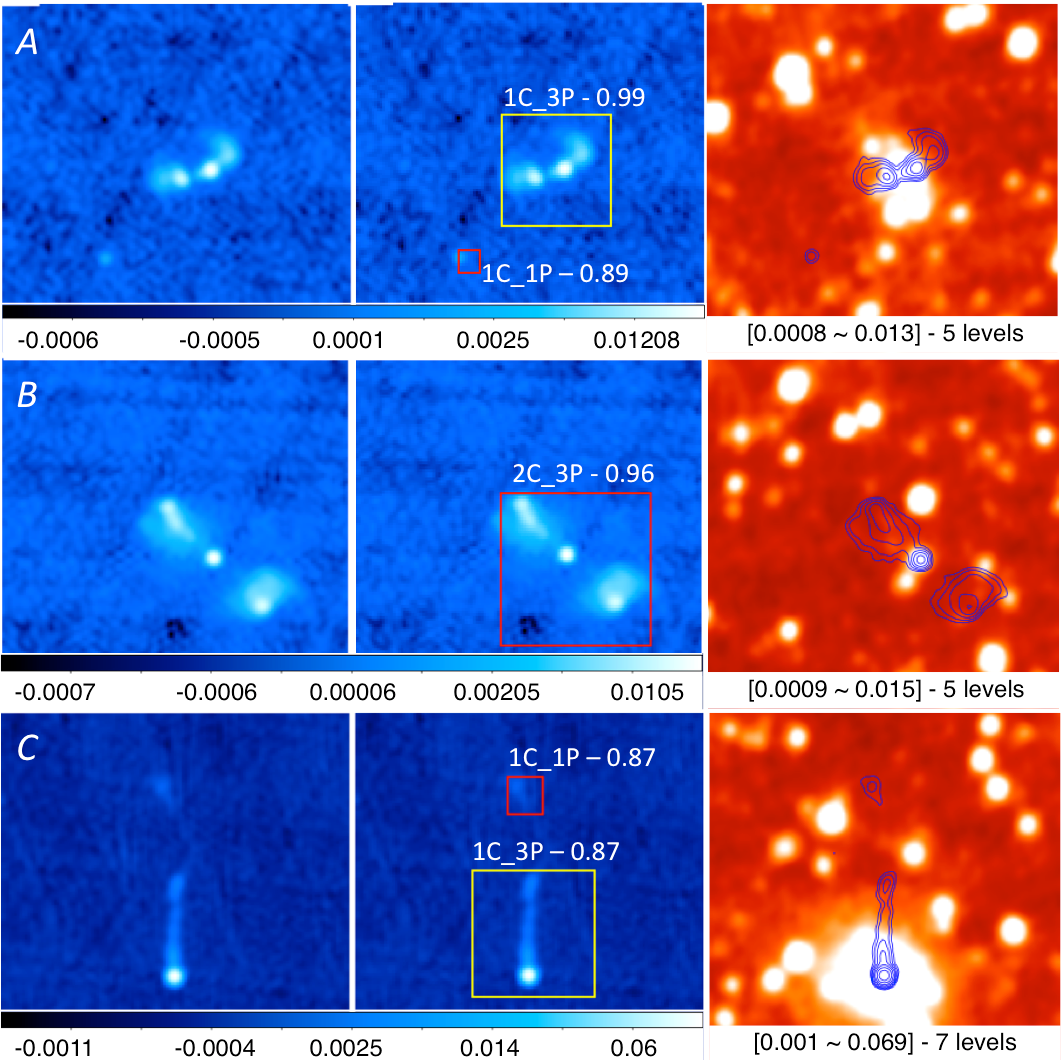}
    \caption{Three classification examples ($A$, $B$, and $C$) on RGZ subjects --- each of them 3' $\times$ 3' in size --- \texttt{FIRST J081700.6+571626}, \texttt{FIRST J070822.2+414905}, and \texttt{FIRST J083915.7+285125}. 
    The first column shows the FIRST radio emission. The second column shows the {\cla} output --- a box encompassing each identified source, and its morphology is labelled as \(i\)\textbf{C}\_\(j\)\textbf{P}, where \(i\) and \(j\) denotes the number of radio components and the number of radio peaks respectively. Each morphology label is associated with a score between 0 and 1, indicating the probability of the quoted morphology class. The first two columns share the same color bar at the bottom, denoting flux density values in Jy/beam. The last column shows the corresponding WISE infrared image overlaid with \(5\sigma \) radio contours. The contour levels (Jy/beam) are shown at the bottom of each infrared image.}
    \label{fig:detection_examples}
\end{figure}

\subsection{Classification examples}
Before discussing the dataset used for this study, we first present some classification examples shown in Figure \ref{fig:detection_examples}. Given a pair of FIRST and WISE images, {\cla} directly outputs the following in approximately 200 milliseconds when measured on a single Tesla K40c GPU with 12GB GPU memory.
\begin{itemize}
\item the location and size of each detected radio source shown as a bounding box predicted by {\cla} during testing,
\item the morphology \(m\) of each detected source labelled as `\(iC\_jP\)', where \(i\) is the number of components, and \(j\) is the number of flux-density peaks, and
\item the probability ($P$-value) of \(m\) for each detected radio source 
\end{itemize}

Following the definitions from the RGZ project \citep{banfield15} and (Wong et al., in preparation), each RGZ \textit{subject} is a 3~arcmin by 3~arcmin FoV inspected by the citizen scientists, and the term \textit{component} refers to discrete individual radio source components identified at the 4-sigma flux-density threshold level, and the term \textit{peak} refers to the number of resolved peaks that are identifiable within each class of objects.  For example, a double-lobed radio galaxy with small angular extent and no radio core may be identified as a source with one component-two peaks (1C-2P) or a two component-two peaks (2C-2P) if the two lobes appear disconnected in the radio image.

In example $A$ of Figure \ref{fig:detection_examples}, {\cla} correctly identifies two radio sources --- the large source has 1 component with 3 peaks, and the small one has 1 component with 1 peak. Both detections are given probabilities (0.99 and 0.89) much higher than 0.8. This example shows {\cla} is able to identify sources at different scales in the same image. In example $B$, {\cla} correctly locates a source with two radio components and three peaks (as per DR1) with a probability of 0.96. This example shows that {\cla} is able to identify extended sources. 

In example $C$, {\cla} detects two independent sources, and assigns the same probability (0.87) to both of them. Although the real radio source is much larger based on the NRAO VLA Sky Survey \citep[NVSS; ][]{condon98}, extending beyond the RGZ subject and including both red and yellow boxes as its internal components, {\cla}'s prediction is still highly plausible considering its view is completely restricted within the 3' by 3' RGZ subject.

It should be noted that all radio and infrared images in Figure \ref{fig:detection_examples} are taken from the testing set (cf. Section \ref{sec:train_test_set}), which {\cla} does not see during training.






\subsection{Consensus level}
\label{sec:cl}
We use two criteria to select fields from DR1 in order to create the training set and the testing set for {\cla}. First, for each selected subject \(f\), we ensure \textit{all} radio sources within \(f\) have a user-weighted Consensus Level (CL) no less than 0.6. CL measures the relative agreement levels of classification among citizen scientists and is defined in \citet{banfield15} as the largest fraction of the total classifications for a radio source that have been agreed upon. 
This is to ensure most radio sources exposed to {\cla} are morphologically human-resolvable. 

Second, 
we ensure \textit{every} radio source within \(f\) has fewer than four components and four peaks. This is because radio sources that (1) have a CL \( \geq 0.6\) and (2) have more than three components or peaks are rare as shown in Table \ref{tab:dataset_all_morph}. Inclusion of these sources into our study leads to highly unbalanced training and testing sets. Although dealing with unbalanced data sets is an on-going machine learning research topic~\citep{he2009learning}, in this paper we focus solely on the main demographic of multi-component/peak sources, and leave for future work the issue of tackling unbalanced datasets with rarer sources.


\begin{table}
	\centering
	\caption{The number of DR1 radio sources (CL \(\geq 0.6\)) for each morphology class. The number of components and peaks for each source in this table is determined by RGZ DR1. Sources with more than 3 components/peaks are rare, and are excluded from this study to avoid unbalanced data sets. Sources with a morphology in the bold face (i.e. 1C-1P, 1C-2P, 1C-3P, 2C-2P, 2C-3P, and 3C-3P) are included in the training and testing sets for this study.}
	\label{tab:dataset_all_morph}
	\begin{tabular}{rr|rr|rr} 
		\hline
		Morph & Count & Morph & Count & Morph & Count\\
		\hline \hline
		\textbf{1C-1P} & 49,766 & 2C-5P & 36 & 4C-6P & 7\\
		\textbf{1C-2P} & 14,242 & 2C-6P & 7 & 4C-7P & 5\\
		\textbf{1C-3P} & 1,412 & 2C-7P & 2 & 5C-5P & 28\\
		1C-4P & 191 & \textbf{3C-3P} & 1,347 & 5C-6P & 11\\
		1C-5P & 28 & 3C-4P & 163 & 5C-7P & 1\\
		1C-6P & 12 & 3C-5P & 20 & 6C-6P & 2\\
		1C-7P & 3 & 3C-6P & 13 & 6C-7P & 1\\
		\textbf{2C-2P} & 9,772 & 3C-7P & 2 & 7C-7P & 2\\
		\textbf{2C-3P} & 1,220 & 4C-4P & 99 & 7C-10P & 1\\
		2C-4P & 181 & 4C-5P & 18 &  & \\
		
		\hline
	\end{tabular}
\end{table}

\begin{figure}
	\includegraphics[width=\columnwidth]{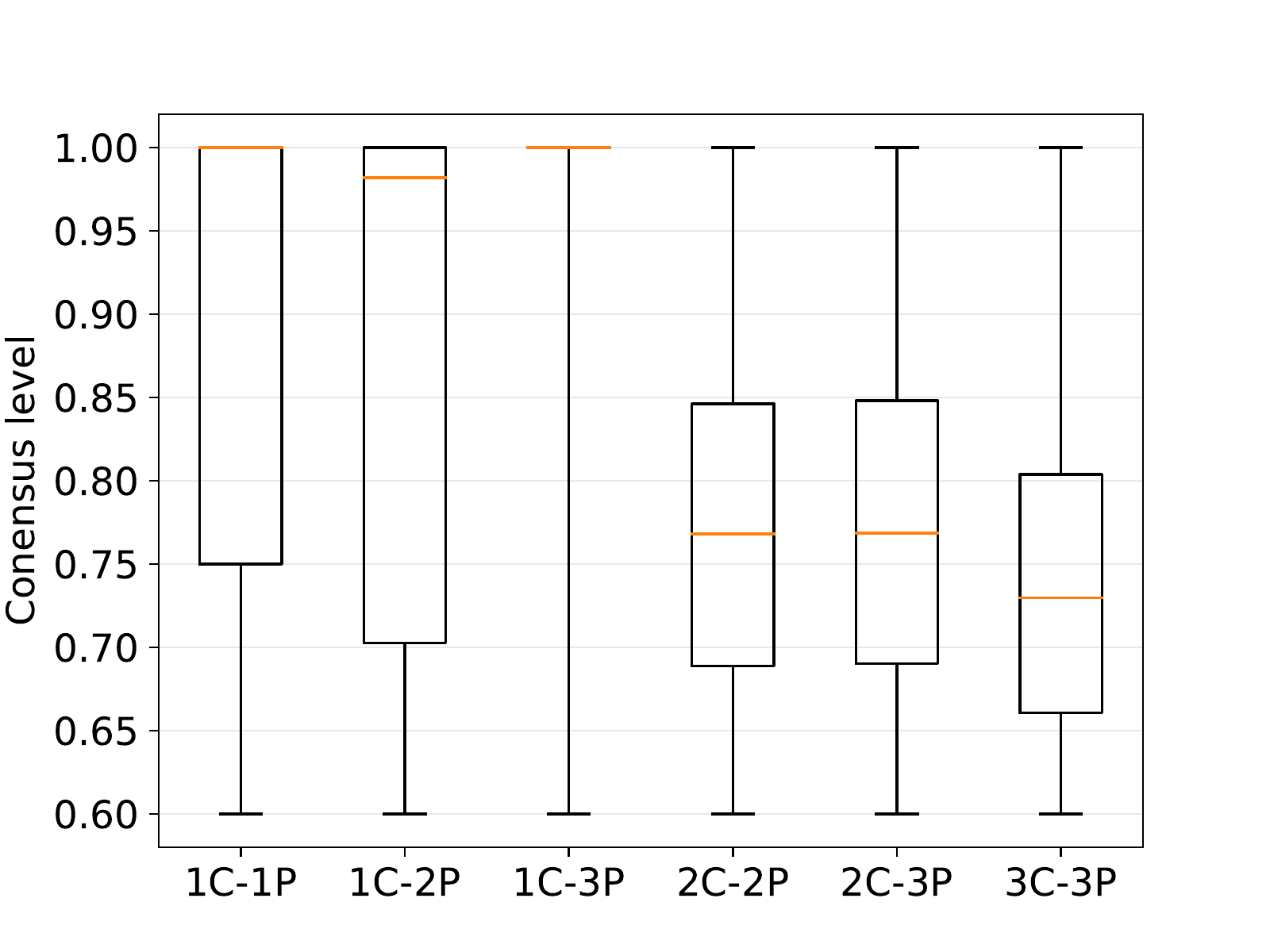}
    \caption{The distribution of the consensus level (CL) across six morphology classes in the data set that consists of 10,744 RGZ subjects selected from DR1. The whiskers above and below the box represent the maximum and minimum CL (fixed at 0.6 by the first criterion). The box itself spans the third and the first quartile CL. Note that since 80\% of 1C-3P sources have a CL of 1.0, its box is reduced to a single horizontal line when its interquarter range becomes 0. The horizontal (orange) line inside each box is the median. 
    }
    \label{fig:cl_dist}
\end{figure}

Upon applying the above two selection criteria on DR1, we obtain a data set $E$ that has 10,744 RGZ subjects.
Figure \ref{fig:cl_dist} shows the CL distribution of sources in $E$ across the six morphology classes. Most one-component sources have high CL (with medians of 1C-1P and 1C-3P reaching the maximum CL value of 1.0) due to their relative simplicity. In particular, 1,133 out of 1,412 (\(80\%\)) 1C-3P sources have CLs equal to 1.0, which explains why its box in Figure \ref{fig:cl_dist} is collapsed to a line when the first and third quartiles are both 1.0. 
1C-2P has a slightly lower median CL (\(0.98\)) than that of 1C-1P or 1C-3P, 
but its third quartile also reaches 1.0. On the other hand, multi-component/peak sources have much lower CLs in general. For example, most CLs of both 2C-2P and 2C-3P are distributed between 0.69 and 0.85 with 0.76 as their medians. CLs of 3C-3P sources have a similar median of 0.73 and a distribution between 0.66 and 0.81. Although CLs vary between these two groups of single-/multi-component sources, 
reaching consensus naturally becomes harder with increasing  morphological complexity associated with multi-component sources. Given the above reasons we define the morphology classes listed in Table \ref{tab:dataset_dist} as \textit{ground-truth morphology} for both training and testing. 

\subsection{Training and testing sets}
\label{sec:train_test_set}
We randomly split the data set $E$ described in Section \ref{sec:cl} into two subsets --- the training set that contains 6,141 subjects, and the testing set that contains 4,603 subjects.
Their basic properties are summarized in Table \ref{tab:dataset_prop}. Table \ref{tab:dataset_dist} shows the morphology distribution of radio sources across six combinations of components and peaks. Although the number of 1C\_1P sources is far greater than sources of other morphology classes in Table \ref{tab:dataset_dist}, the evaluation in Section \ref{sec:evaluation} will show that {\cla} is not biased towards 1C\_1P sources.

\begin{table}
	\centering
	\caption{Basic properties of the training and testing data sets used by {\cla}. One subject may contain multiple sources. One source may contain multiple components and multiple radio peaks.}
	\label{tab:dataset_prop}
	\begin{tabular}{lrrrr} 
		\hline
		Set & Subjects & Sources & Components & Peaks\\
		\hline \hline
		Training & 6,141 & 6,978 & 9,566 & 12,441\\
		Testing & 4,603 & 4,858 & 7,397 & 9,885\\
		\hline
		Total & 10,744 &11,836 & 16,963 & 22,326\\
		\hline
	\end{tabular}
\end{table}

\begin{table}
	\centering
	\caption{Number of radio sources (Consensus Level \(\geq 0.6\)) for each morphology class in the training and testing data sets. A morphology class is represented as a combination of the number of Components \(C\) and  the number of Peaks \(P\). Consensus level is discussed in Section \ref{sec:cl} and further illustrated in Table \ref{tab:dataset_all_morph}}
	\label{tab:dataset_dist}
	\begin{tabular}{rrrrrrr} 
		\hline
		Set & 1C-1P & 1C-2P & 1C-3P & 2C-2P & 2C-3P & 3C-3P\\
		\hline \hline
		Training & 3,518 & 810 & 728 & 647 & 609 & 666\\
		Testing & 1,782 & 521 & 684 & 604 & 599 & 668\\
		\hline
		Total & 5,300 & 1,331 & 1,412 & 1,251 & 1,208 & 1,334\\
		\hline
	\end{tabular}
\end{table}

To generate the \textit{ground-truth location} --- both \textit{location} and \textit{size} of each known source within a given subject --- we produce a square bounding box for each source based on its physical attributes defined in the RGZ dataset. We use its central location \texttt{RA} and \texttt{DEC} as the box center, and calculate the sky coordinates \(S_c\) of the box's four corners using the RGZ DR1 \texttt{max\_angular\_extent} parameter, which is an estimate of the source's angular size for all RGZ consensus sources as detailed in ~\citet{banfield15} and Wong et al. (in preparation). 
We then convert \(S_c\) into pixel coordinates \(P_c\) that can be processed by imaging software libraries. An extra step is taken to ensure the first element of \(P_c\) represents the top left corner as required by formats such as PNG or JPEG rather than the bottom left corner as in the FITS format. 

\begin{figure}
	\includegraphics[width=\columnwidth]{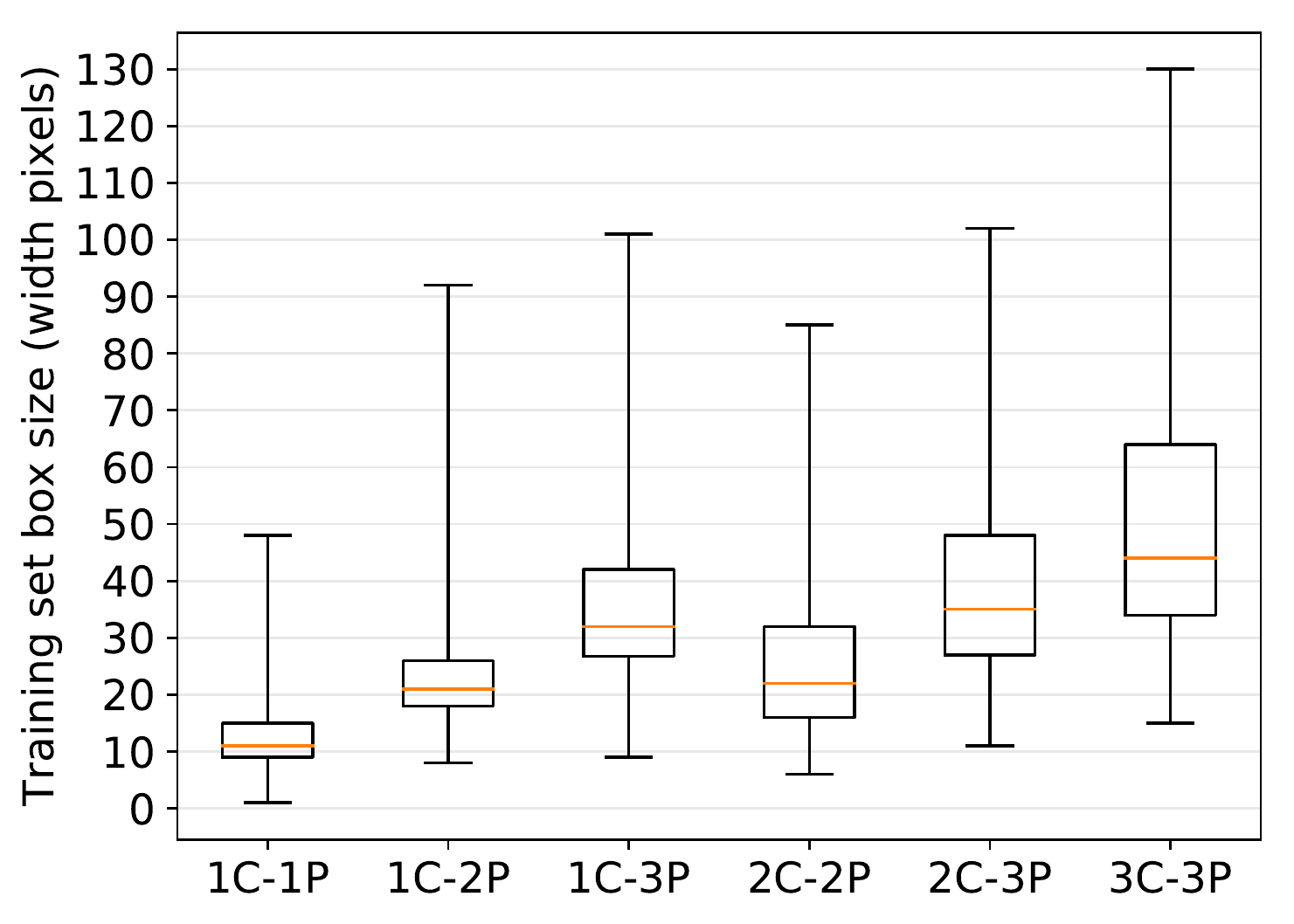}
    \caption{The distribution of bounding box sizes (width or height) in the training set for each morphology class. Note that the FIRST image pixel size is 1.375", therefore the 3' $\times$ 3' angular size of each subject corresponds to \(132 \times 132\) pixels, which sets the maximum possible value of the box size.}
    \label{fig:box_size}
\end{figure}

Figure \ref{fig:box_size} shows the size distribution of generated ground-truth boxes (i.e. radio sources) in the training set. The median size of the box appears positively correlated with the number of peaks, and if two sources have the same number of peaks, the one with more components has a slightly bigger size. Several extraordinarily large three-component sources almost cover the entire image. 

\subsection{Derived datasets}
\label{sec:derived_datasets}
\begin{figure}
	\includegraphics[width=\columnwidth]{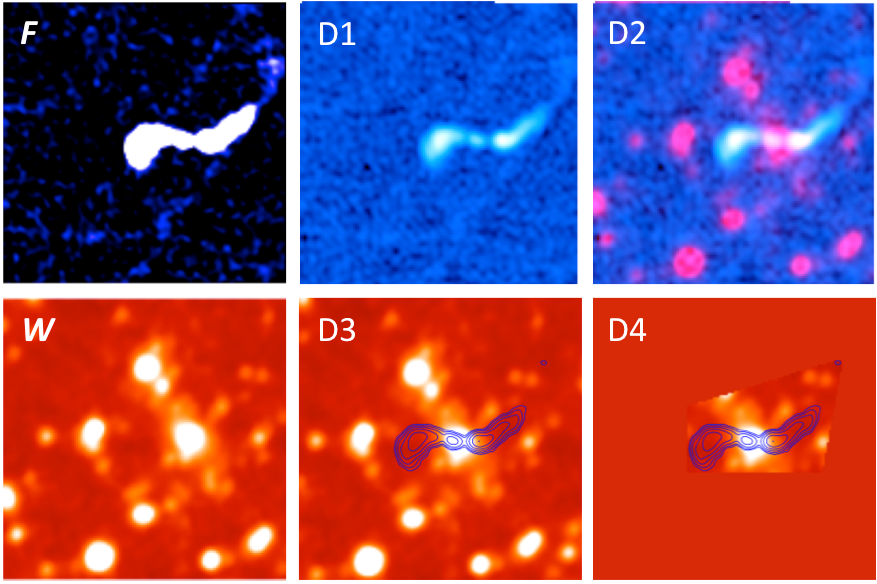}
    \caption{Based on the input FIRST image \texttt{FIRST J014110.8+121353}, examples of derived datasets are shown as D1, D2, D3, and D4. These maps are discussed in more detail in Section~\ref{sec:derived_datasets}.} 
    \label{fig:data_prep}
\end{figure}

The original RGZ dataset contains FIRST radio images (in both FITS and PNG formats) and WISE infrared PNG images. While the beam size of the FIRST survey is 5 arcsec, the size of each FITS pixel is about 1.375 arcsec. Therefore the angular size of a 132 by 132 pixel RGZ subject is $\sim$ 3-arcmin $\times$ 3-arcmin. An example RGZ subject with the radio source \texttt{FIRST J014110.8+121353} is shown as a PNG image $F$ in Figure \ref{fig:data_prep}, and its WISE infrared counterpart is shown as image $W$ underneath $F$. Note that $F$ is exported from the original FITS format as a three-channel (RGB) image under the `cool' colormap using DS9~\citep{joye2013saoimage}. To effectively train {\cla}, we derive four additional datasets --- D1, D2, D3 and D4 --- from \(F\) and \(W\). While both \(F\) and D1 display radio emission only, \(F\) uses the \texttt{DS9 linear-zscale} scale to represent flux values in the PNG format, whereas D1 uses the \texttt{DS9 log-min-max} scale. The rationale of creating D1 is to reveal the internal structures, but potentially at the cost of exposing more background noise. In this example, three separate radio peaks can be identified in D1 by eye but they appear blended together in \(F\). It should be noted that training and testing on datasets F or D1 do not involve any infrared images.

Similar to D1, D2 also uses the \texttt{DS9 log-min-max} scale. However, it increases the intensity of D1's red channel by corresponding pixel values in \(W\) while keeping D1's blue and green channels unchanged. This essentially overlays infrared sources as red blobs on top of radio sources. The intention is to let {\cla} learn interaction patterns between the host galaxy (if detected in WISE) and its surrounding radio emission. D3 aims to achieve the same goal but operates in the opposite direction. It generates \(5\sigma\) contours\footnote{Unlike the RGZ Web interface which uses \(4\sigma\) contours, we selected \(5\sigma\) to reduce potential contamination from noise artefacts that are present in some fields.}  based on surface brightness as recorded in the FIRST FITS file, and then overlays the radio contours on top of \(W\). The RGZ Web user interface allows citizen scientists to transition between $F$ and D3 (with a different level of sigma and contour colors) via a slider. Detailed descriptions of the RGZ interface can be found in \citet{banfield15}.

We notice that there are numerous infrared sources in \(W\) that are not directly related to the overlaid radio contours/sources. Their existence may mislead {\cla} to learn patterns from noise rather than features. To alleviate this issue, D4 generates a convex hull\footnote{http://mathworld.wolfram.com/ConvexHull.html} over (sample points on) all radio contours in D3. The convex hull here denotes the union area enclosed by all radio contours on the infrared image. For each channel \(c\), D4 masks pixels outside the convex hull with the mean pixel value of \(c\) over all images in the training set. As a result, we remove all the infrared signals that do not fall within the convex hull. Since the convex hull covers all radio contours, it should expose sufficient infrared signals to capture the interplay between all radio sources/components. However, this cannot deal with certain special cases where a host galaxy is situated outside the union area formed by all radio source components within a subject. Such examples include remnant radio galaxies (there is no core) or there are faint, compact, separate (i.e. disconnected) lobes on opposite sides of WISE objects in the RGZ subjects. For these cases D3 is perhaps more appropriate. Future research should investigate more optimal and generalizable data fusion techniques that, for example, have the advantages of both D3 and D4.

\section{Data pipeline}
\label{sec:data_pipeline}
\begin{figure}
	\includegraphics[width=\columnwidth]{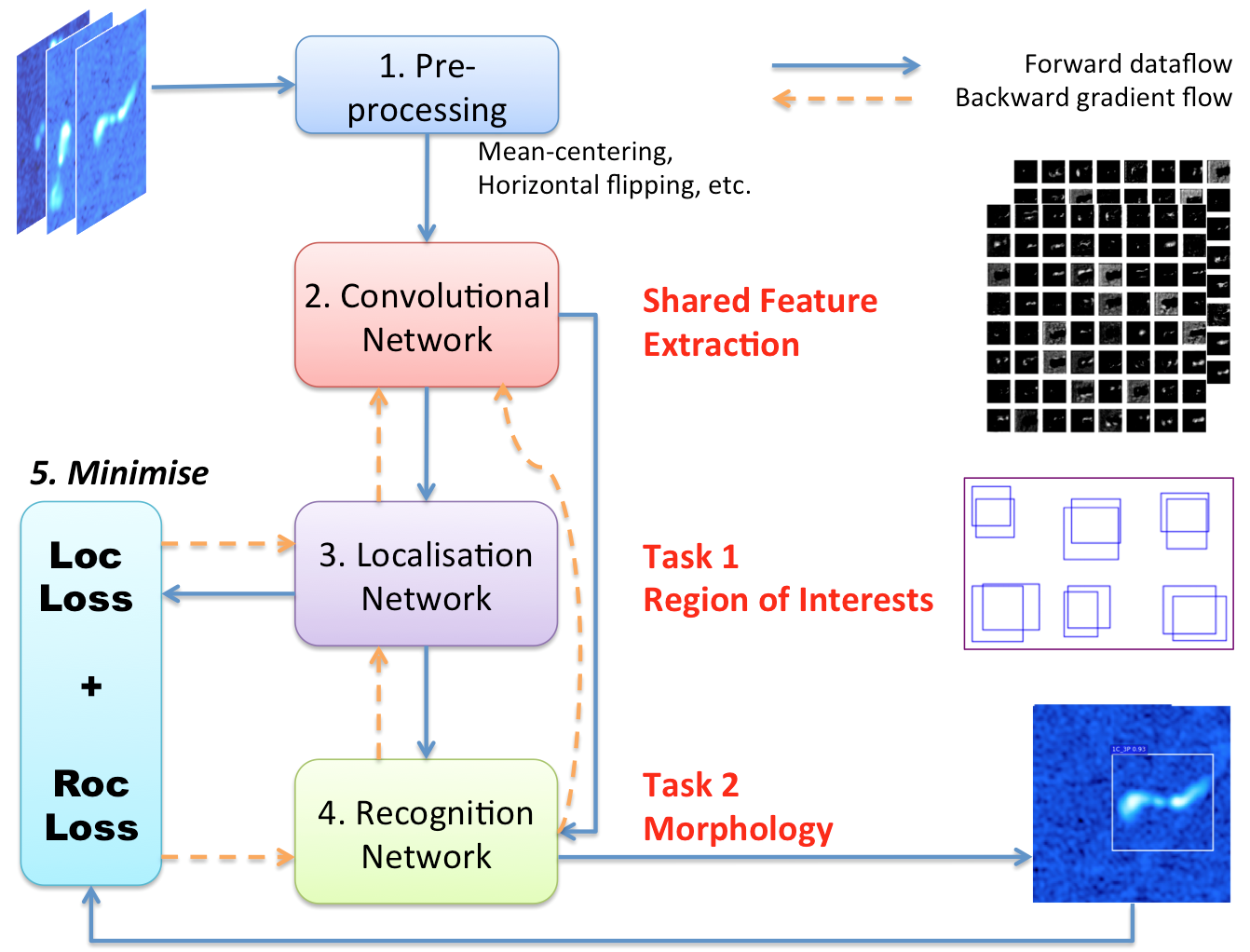}
    \caption{The end-to-end training pipeline that learns two related tasks simultaneously. The solid arrow denotes forward dataflow, in which a list \(L_f\) of parameterised functions are computed consecutively on each image batch. The output from \(L_f\), known as `prediction', is fed to the loss function (step 5) to calculate the error between ground truths and predictions. The error is converted to the global gradient, and propagated (via local gradient updates) backward to each function in \(L_f\) so that they can adjust their parameters to reduce the errors. The alteration of forward dataflow and backward gradient flow is repeated for each image batch, iteratively minimizing the loss function until the error converges below a threshold.}
    \label{fig:pipeline_concepts}
\end{figure}

In this section, we introduce our dual-task, end-to-end data pipeline based on the Faster R-CNN method. By dual-task, we mean the pipeline trains a detector to learn two separate tasks --- \textit{localization} and \textit{recognition}. While both tasks share the same input features derived from the convolutional layer, the learning outcome of the first task will directly affect the learning performance of the second task. 
By end-to-end, we mean the entire training pipeline has only a single step of optimization, and the two tasks are trained simultaneously in a single training iteration. It also means little human involvement is needed for deriving hand-crafted features, and feature extraction is driven primarily by convolutional kernels learned from training sets rather than prior assumptions imposed by experts.
Figure \ref{fig:pipeline_concepts} shows the data pipeline during the training stage, which we explain in detail below.

\subsection{Pre-processing}
\label{sec:pre-processing}
 In the first phase, three pre-processing operations --- zero-centering, size scaling, and horizontal flipping --- are performed on-the-fly in a streaming mode on each input image.
 
 Zero-centering involves (1) calculating the mean \(\mu_{C}\) for each channel \(C\) across the entire training set, and (2) subtracting \(\mu_{C}\) from each pixel of \(C\) in a given input image \(I\). Since the subsequent convolutional filters are also initialized as truncated Gaussians centered at zero with a small standard deviation (0.01 in our training pipeline), filter response \(R\) from \(I\) is also zero-centered with a small variance. \(R\) is then transformed by the subsequent Rectified Linear Unit (ReLU)~\citep{nair2010rectified} activation function defined as \(A(x) = \max(0, x)\) to output the feature map.  
 It has been reported~\citep{krizhevsky2012imagenet} that ReLU, while simple and efficient to compute, accelerates the convergence of the optimization procedure such as the Stochastic Gradient Descent (SGD) by a factor of six. Moreover, it often results in superior solutions~\citep{glorot2011deep, zeiler2013rectified} than more traditional, sigmoid-like activation functions. During SGD, if all \(R\)s are closely centered around zero, given a fixed pixel \(p\), it is highly likely \(p\) in some \(R\) becomes positive to activate ReLU (for non-zero gradient descent), which will be less likely without zero-centering.

The largest receptive field\footnote{Section \ref{sec:conv_net} describes the concept of \textit{receptive field} and Eq. \ref{eq:receptive_field} defines its calculation} of a neuron in the last shared convolutional feature map is 228. Figure \ref{fig:box_size} shows that the median box size of 3C\_3P is slightly below 50. Therefore we scale up the image size by a factor of \(228/50 = 4.56\) to match the median box size to the final receptive field size. This involves increasing the height and width of the (fused) image from \(132 \times 132\) pixels to \(600 \times 600\) pixels using the bilinear interpolation. Moreover, we scale up coordinates of each ground-truth box by the same factor 4.56. It should be noted that scaling up the image size does not scale pixel intensities, which is a useful pre-processing technique~\citep{stark18} that we will explore in our future work for {\cla}.

During training, we use horizontal flipping to create a symmetric counterpart for a given input image \(I\) by appending an extra image \(I'\) that reverses the pixels order along the horizontal axis of \(I\). This allows {\cla} to expect different source orientations other than provided in the original training set. We also create horizontally flipped ground-truth boxes to match the flipped image \(I'\).
 
 \subsection{Convolutional network}
 \label{sec:conv_net}
The Convolutional Network (\textbf{ConvNet}) --- including Layers 1 to 17 in Table \ref{tab:network_architecture} --- performs feature extraction in order to produce feature maps shared by both tasks and their associated networks. The basic two dimensional convolution operation at each layer can be expressed as:
\begin{equation}
\label{eq:convnet}
Y(m, n) = A\Big(\sum_{k=1}^{C}\sum_{j=1}^{W}\sum_{i=1}^{H}X(k, i, j)K(k, m - i, n - j) + B\Big)
\end{equation}
In Eq.\ref{eq:convnet} \(X\) is an input image or an intermediary tensor with \(C\) planes (or `channels' for RGB images), height \(H\), and width \(W\). \(Y\) is the output of the convolution, i.e. the \textbf{feature map}. \(Y(m, n)\) denotes \(Y\)'s value at row \(m\) and column \(n\). \(K\) is a centre-originated kernel with channel \(C\), height and width \(s\),  
and \(K(k, a, b) = 0\) if \(|a|\) or \(|b| > \frac{s}{2}\).  Note that a feature map of one convolutional layer becomes the input (i.e. \(X\)) of the next convolutional layer. \(B\) is the \textit{bias} tensor that has the same dimensions as the feature map \(Y\). \(A\) is the element-wise ReLU activation function. Only \(K\) and \(B\) have learnable parameters that are updated during back-propagation through SGD.

We use the first 17 layers (13 weight layers and 4 pooling layers) from the VGG-16 (Configuration D) network \citep{simonyan2015} as the architecture of the convolutional network. This is shown in Table \ref{tab:network_architecture} from layer 1 to layer 17). Compared to other convolutional networks, a neuron in a VGG-16 convolution feature map has a smaller local field of view --- the \textit{receptive field}~\citep{hubel1962receptive} --- a \(3 \times 3\) region from its input layer. However, stacking multiple convolutional layers gradually increases the global receptive field --- i.e. the region in the input image. Neurons in the final feature map (i.e. layer 17 in Table \ref{tab:network_architecture}) has a receptive field of size \(228 \times 228\) when \(k\) is set to 17 in Eq. \ref{eq:receptive_field}:
\begin{equation}
\label{eq:receptive_field}
r_{k} = r_{k - 1} + [(f_{k} - 1) \times \prod_{i=1}^{k - 1} s_{i}]
\end{equation}
where \(r_{k - 1}\) denotes the size of the receptive field of neurons at layer \(k - 1\), \(f_{k}\) is the filter width/height (the third column of Table \ref{tab:network_architecture}) at layer \(k\), and \(s_i\) is the stride of layer \(i\) (the fourth column of Table \ref{tab:network_architecture}).
More importantly, stacking increases the number of non-linear activations since each convolutional layer has its own ReLU non-linearities. It is these non-linearities that ultimately offer the network discriminative capabilities for feature extraction. It should be noted that the size $S$ of the receptive field of a single neuron does not limit {\cla} from detecting sources larger than $S$. This is because a feature map consists of multiple neurons, which collectively can detect much larger objects. 

\begin{figure}
	\includegraphics[width=\columnwidth]{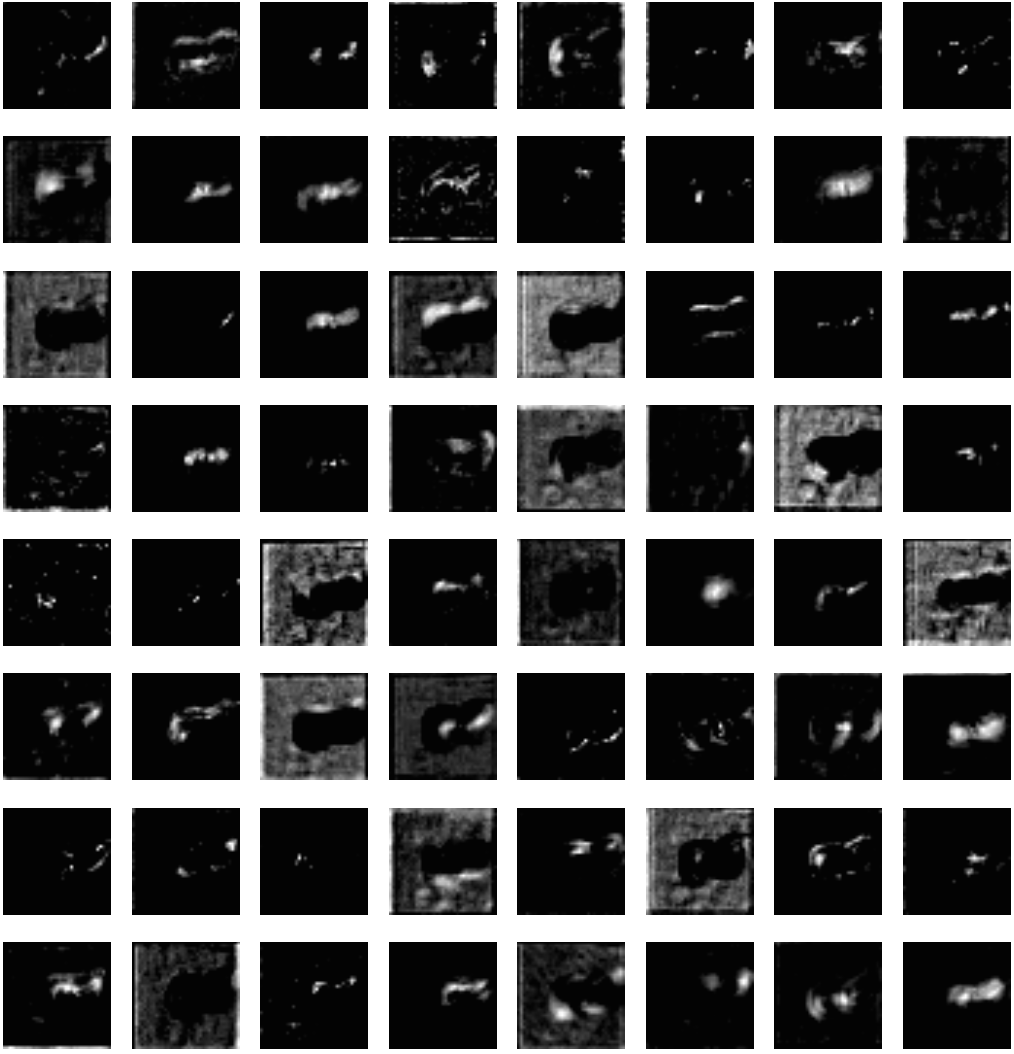}
    \caption{64 feature maps produced by Layer 18 (Table \ref{tab:network_architecture}) given the input image \texttt{FIRST J014110.8+121353} (i.e. the example D1 dataset in Figure \ref{fig:data_prep}). The first 64 of the total 512 channels are shown, and each channel is visualized as a (37 by 37) grey-scale matrix, and white colors denote matrix elements with higher values. The feature map at row 3 column 5 (R3C5, channel 21) appears to be the outcome of cutting out the entire radio emission, revealing the overall contour of the source. 
    R2C1 and R2C3 (channel 9 and 11) appear to represent the top and bottom part of the source respectively as if they were separated by a gap tilting along the direction of the jet. More interestingly and importantly, we always find similar features at the same channel for different input images. This shows that the convolution kernels have learned something intrinsic and constant across different subjects. }
    \label{fig:convnet_output}
\end{figure}

Figure \ref{fig:convnet_output} shows feature maps produced by the last shared convolution network layer (i.e. layer 17 \textit{Conv5\_3} in Table \ref{tab:network_architecture}). The features are extracted from the input image \texttt{FIRST J014110.8+121353} shown in Figure \ref{fig:data_prep}. The extraction were performed after the completion of training, which consists of 80,000 iterations of forward computation and backward propagation in order to find optimal values for all the kernel weights in the ConvNet. Visual inspection reveals some resemblance between the input image and each one of the 64 feature maps that capture the shape of the radio jets. However, each feature map exposes distinct features produced by a different set of kernels, each of which has learned to find and match a unique set of patterns from its input tensors. Collectively, these feature maps provide input for the two tasks to learn.

The parameters in the 13 weight layers are essentially shared by all following layers starting from layer 18, and are learned jointly by both task 1 (\textit{localization}) and task 2 (\textit{recognition}). To initialize weights in layer 1 to 17, we load public VGG-16 model weights\footnote{\url{http://www.deeplearningmodel.net/}} pre-trained from the ImageNet~\citep{ILSVRC15}. 
We then freeze the weights of the first four convolutional layers (1, 2, 4, and 5) by assuming low-level features learned by these filters remain constant across different domains, and set free weights in higher layers in order for them to learn higher level structures and patterns unique to radio galaxy morphology. We choose these four layers because their neurons have relatively small receptive fields --- 5 \(\times\) 5, 6 \(\times\) 6, 14 \(\times\) 14, and 16 \(\times\) 16 pixels on the scaled 600 \(\times\) 600 pixel image --- well suited to capture low-level, local features\footnote{The next convolutional layer 7 has a receptive field of 32 \(\times\) 32 pixels (thus 7 \(\times\) 7 pixels on the original image), which equate to the first quartile size of 1C\_1P sources, and therefore too large for low-level feature extraction.}. Compared to learning these weights from scratch, we find that using pre-trained weights significantly improves the detection performance given the same amount of training time.

\begin{table*}
\begin{center}
\caption{The Faster R-CNN model used by {\cla}, which consists of three networks --- the Convolutional network (\textit{ConvNet}, layers 1 to 17), the Localization network (\textit{LocNet}, layers 18 to 22), and the Recognition network (\textit{RecNet}, layers 23 to 29). Functions in \textit{ConvNet} are either convolution operations (e.g. \texttt{Conv1\_1}) or max pooling operations (e.g. \texttt{MaxPool\_1}). Functions in \textit{LocNet} and \textit{RecNet} are explained in Section \ref{sec:loc_net} and Section \ref{sec:rec_net} respectively. The Filter/Input tensor size in \textit{ConvNet} refers to the convolutional/pooling filter size, whose first three dimensions (left to right) are the height, width, and depth of the filter.
For convolutional filters, the fourth dimension denotes the number of filters. The Output tensor size in \textit{ConvNet} refers to the height, width, and depth of the output feature map. Convolution operations in \textit{LocNet} --- \texttt{RPN\_Conv}, \texttt{Anchor\_Cls\_Conv}, and \texttt{Anchor\_Reg\_Conv} --- also have the same four-dimension filter sizes. Input and output tensor sizes for other functions are explained in Section \ref{sec:loc_net} and \ref{sec:rec_net}. All Activations associated with convolution and dense-layer functions (i.e. \texttt{FC\_6} and \texttt{FC\_7}) are ReLU. The model in total consists of \(136,777,443\) ``trainable" parameters, which are summed over all rows of the last column.}
  \label{tab:network_architecture}
  \begin{tabular}{l | l | l | c | c | l | r |  }
    \hline
    Layer & Function & Filter/Input tensor size & Activation & Stride & Output tensor size & No. of parameters \\ \hline
    \hline
    0 & Input & 600 $\times$ 600 $\times$ 3 & - & - & - & 0 \\ 
    1 & Conv1\_1 & 3 $\times$ 3 $ \times$ 3 $\times$ 64 & ReLU & 1 & 600 $\times$ 600 $\times$ 64 & 1,728 \\
    2 & Conv1\_2 & 3 $\times$ 3 $ \times$ 64 $\times$ 64 & ReLU &  1 & 600 $\times$ 600 $\times$ 64 & 36,864 \\
    3 & MaxPool\_1 & 2 $\times$ 2 $\times$ 64 & - &  2 & 300 $\times$ 300 $\times$ 64 & 0 \\
    4 & Conv2\_1 & 3 $\times$ 3 $ \times$ 64 $\times$ 128 & ReLU & 1 & 300 $\times$ 300 $\times$ 128 & 73,728 \\
    5 & Conv2\_2 & 3 $\times$ 3 $ \times$ 128 $\times$ 128 & ReLU & 1 & 300 $\times$ 300 $\times$ 128 & 147,456 \\
    6 & MaxPool\_2 & 2 $\times$ 2 $\times$ 128 & - &  2 & 150 $\times$ 150 $\times$ 128 & 0 \\
    7 & Conv3\_1 & 3 $\times$ 3 $ \times$ 128 $\times$ 256 & ReLU &  1 & 150 $\times$ 150 $\times$ 256 & 294,912 \\
    8 & Conv3\_2 & 3 $\times$ 3 $ \times$ 256 $\times$ 256 & ReLU &  1 & 150 $\times$ 150 $\times$ 256 & 589,824 \\
    9 & Conv3\_3 & 3 $\times$ 3 $ \times$ 256 $\times$ 256 & ReLU &  1 & 150 $\times$ 150 $\times$ 256 & 589,824 \\
    10 & MaxPool\_3 & 2 $\times$ 2 $\times$ 256 & - &  2 & 75 $\times$ 75 $\times$ 256 & 0 \\
    11 & Conv4\_1 & 3 $\times$ 3 $ \times$ 256 $\times$ 512 & ReLU &  1 & 75 $\times$ 75 $\times$ 512 & 1,179,648 \\
    12 & Conv4\_2 & 3 $\times$ 3 $ \times$ 512 $\times$ 512 & ReLU &  1 & 75 $\times$ 75 $\times$ 512 & 2,359,296 \\
    13 & Conv4\_3 & 3 $\times$ 3 $ \times$ 512 $\times$ 512 & ReLU &  1 & 75 $\times$ 75 $\times$ 512 & 2,359,296 \\
    14 & MaxPool\_4 & 2 $\times$ 2 $\times$ 512 & - &  2 & 37 $\times$ 37 $\times$ 512 & 0 \\
    15 & Conv5\_1 & 3 $\times$ 3 $ \times$ 512 $\times$ 512 & ReLU &  1 & 37 $\times$ 37 $\times$ 512 & 2,359,296 \\
    16 & Conv5\_2 & 3 $\times$ 3 $ \times$ 512 $\times$ 512 & ReLU &  1 & 37 $\times$ 37 $\times$ 512 & 2,359,296 \\
    17 & Conv5\_3 & 3 $\times$ 3 $ \times$ 512 $\times$ 512 & ReLU &  1 & 37 $\times$ 37 $\times$ 512 & 2,359,296 \\
    \hline
    18 & RPN\_Conv & 3 $\times$ 3 $ \times$ 512 $\times$ 512 & ReLU &  1 & 512 $\times$ 37 $\times$ 37 & 2,359,296 \\
    19 & Anchor\_Cls\_Conv & 1 $\times$ 1 $ \times$ 512 $\times$ 12 & - &  1 & 12 $\times$ 37 $\times$ 37 & 6,144 \\
     & Anchor\_Cls\_Conv\_RS & 12 $\times$ 37 $\times$ 37 & - &  - & (6 $\times$ 37) $\times$ 37 $\times$ 2 & 0 \\
    20 & Anchor\_Cls\_Softmax & (6 $\times$ 37) $\times$ 37 $\times$ 2 & - &  - & (6 $\times$ 37) $\times$ 37 $\times$ 2 & 0 \\
     & Anchor\_Cls\_Softmax\_RS & (6 $\times$ 37) $\times$ 37 $\times$ 2 & - &  - & 37 $\times$ 37 $\times$ 12 & 0 \\
    20 & Anchor\_Target & $12 \times 37^2$, gt\_box $\times$ 5 & - & - & $37^2 \times$ 12,  $37^2 \times$ 24 & 0 \\
    19 & Anchor\_Reg\_Conv & 1 $\times$ 1 $ \times$ 512 $\times$ 24 & - &  1 & 24 $\times$ 37 $\times$ 37 $$ & 12,288 \\
    21 & RoI\_Proposal & $37^2 \times$ 12, $24 \times 37^2$ & - & - & NMS\_TopN $\times$ (4 + 1) & 0 \\
    22 & RoI\_Proposal\_Target & NMS\_TopN $\times$ 5, gt\_box $\times$ 5 & - & - & RoI\_batch $\times$ 1, RoI\_batch $\times$ 28 & 0 \\
    \hline
    23 & ST\_RoI\_Pool & 37 $\times$ 37 $\times$ 512, RoI\_batch $\times$ 5 & - & - & RoI\_batch $\times$ 7 $\times$ 7 $\times$ 512 & 0 \\
    24 & FC\_6 & RoI\_batch $\times$ 7 $ \times$ 7 $\times$ 512 & ReLU &  - & RoI\_batch $\times$ 4096 & 102,764,544 \\ 
    25 & DropOut\_6 & RoI\_batch $\times$ 4096 & - &  - & RoI\_batch $\times$ 4096 & 0 \\ 
    26 & FC\_7 & RoI\_batch $\times$ 4096 & ReLU &  - & RoI\_batch $\times$ 4096 & 16,781,312 \\ 
    27 & DropOut\_7 & RoI\_batch $\times$ 4096 & - &  - & RoI\_batch $\times$ 4096 & 0 \\ 
    28 & FC\_Cls\_Score & RoI\_batch $\times$ 4096 & - &  - & RoI\_batch $\times$ 7 & 28,679 \\ 
    28 & FC\_Reg\_Pred & RoI\_batch $\times$ 4096 & - &  - & RoI\_batch $\times$ (7 $\times$ 4) & 114,716 \\ 
    29 & Cls\_SoftMax & RoI\_batch $\times$ 7 & - &  - & RoI\_batch $\times$ 7 & 0 \\ 
    \hline
  \end{tabular}
\end{center}
\end{table*}

\begin{figure}
	\includegraphics[width=\columnwidth]{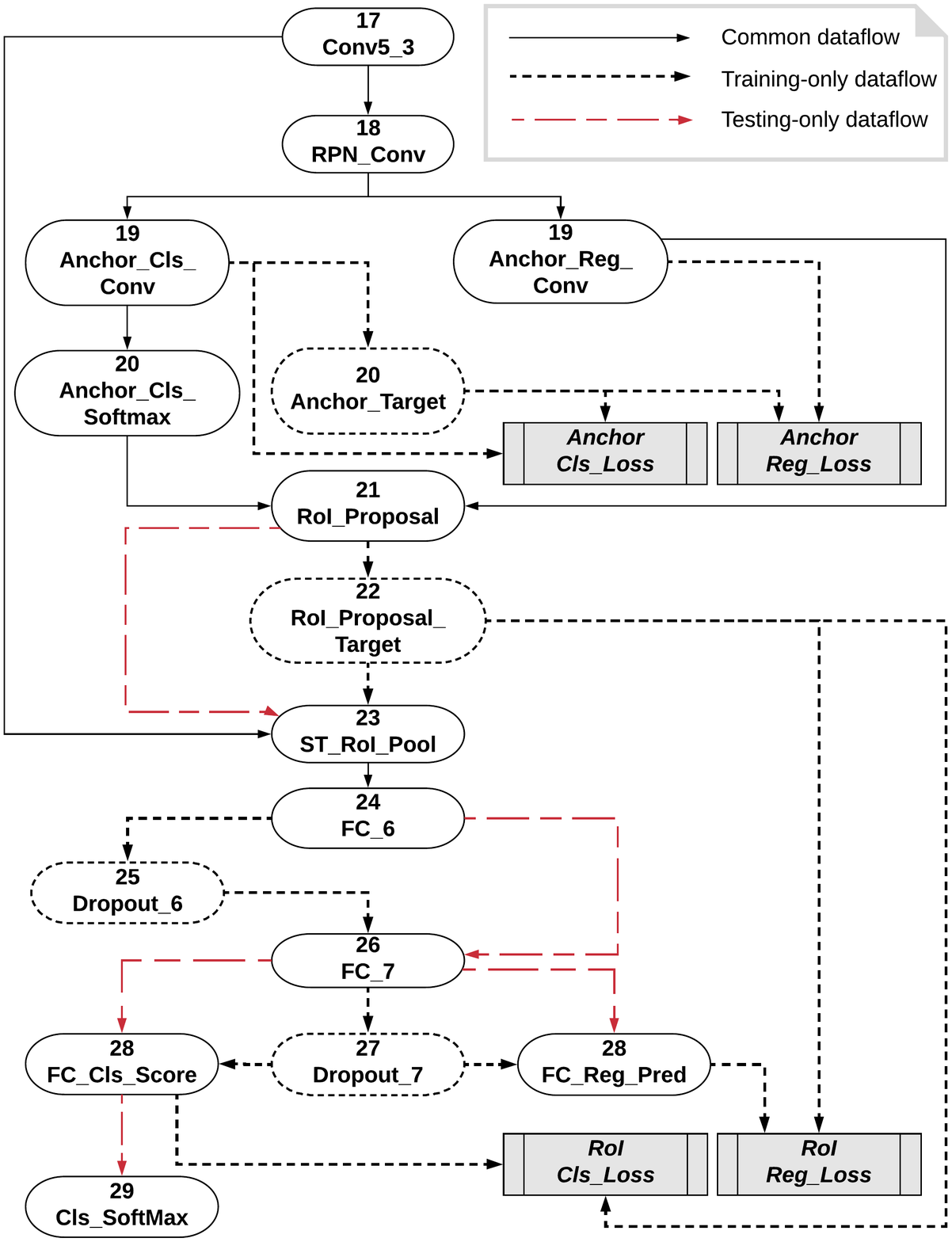}
    \caption{A dataflow diagram for \texttt{LocNet} and \texttt{RecNet}. Each ellipse represents a \texttt{Function} defined in the second column of Table \ref{tab:network_architecture}. Solid ellipses appear in both training and testing, but dotted ones are used for training only. For example, 
    \texttt{Anchor\_Target} and \texttt{RoI\_Proposal\_Target} dynamically generate ground truths for training given a subject --- i.e. positive and negative anchors in \texttt{Anchor\_Target} or proposal-source offsets and morphology class for each proposal in \texttt{RoI\_Proposal\_Target}. These two operations are only used during training, and are removed during testing. Similarly, solid arrows, which denote the dataflow between two data transformations, appear in both training and testing, and dotted ones are used only for training, and dashed ones represent dataflows for testing only. The four gray rectangles denote the four loss functions --- Eq. \ref{eq:loss_anch_cls}, Eq. \ref{eq:loss_anch_reg}, Eq. \ref{eq:loss_roi_reg}, Eq. \ref{eq:loss_roi_cls}  --- in a clock-wise order. Since loss functions are minimised during training, dataflows that provide inputs to these functions are all dotted arrows.}
    \label{fig:loc_rec_net}
\end{figure}

\subsection{Localization network}
\label{sec:loc_net}

The localization network (\texttt{LocNet}) --- layers 18 to 22 in Table \ref{tab:network_architecture} --- is trained to propose a set \(\boldsymbol{R}\) of Region of Interest (RoI) proposals (boxes) given a subject, and each RoI proposal \(r \in \boldsymbol{R}\) represents a potential radio source.  
\subsubsection{Regional Proposal Network}
\label{sec:rpn}
The \texttt{LocNet} starts with a mini-network --- the Regional Proposal Network (RPN) ~\citep{ren2017faster}, which consists of two layers of three convolutional functions. Layer 18 
slides 512 filters over Layer 17 \texttt{Conv5\_3}. Each filter outputs a \([37 \times 37]\) matrix, and all filters in total produce a \([512 \times 37 \times 37]\) feature map --- \texttt{RPN\_Conv}. Reshaping it to \([37 \times 37 \times 512]\),
we treat \texttt{RPN\_Conv} as a grid of \(37 \times 37\) pixels, and each pixel \(x_i\) (where \(i = 1 ... 37^2\)) has 512 values. 

The first step of the RPN is to construct \(k\) \textit{anchors}, which are boxes of different sizes and aspect ratios affixed at the centre of each \(x_i\). These \(k\) anchors act as `prior boxes', some of which have the potential to grow into RoI proposals. Since anchors are stationary and input-invariant, they constitute a fixed reference grid to locate radio source candidates across the entire feature map in parallel. All that is left to figure out is \textit{which} and \textit{how} anchors could be shifted and scaled in order to become RoI proposals.

We set \(k = 6\) to cover scales \([1, 2, 4, 8, 16, 32]\) and aspect ratio \([1.0]\). Since the total number of \textit{strides} on \texttt{Conv5\_3} after four layers of \(2 \times 2\) \textit{max poolings}\footnote{\textit{Stride} controls the offset by which the convolutional filter shifts across the input tensor, and \textit{max pooling} downsamples the input tensor by selecting the maximum pixel in every sub-region convolved with the pooling filter.} (i.e. Layer 3, 6, 10, and 14 in Table \ref{tab:network_architecture}) is \(2^4 = 16\), the anchor sizes projected back on the \(600 \times 600 \) subject are \([16, 32, 64, 128, 256, 512]\).
We keep the anchor aspect ratio to 1 since all ground-truth boxes are squares although the proposed RoI may not be fully square due to the spatial offset described later.
As a result, \texttt{RPN\_Conv} corresponds to a set \(\boldsymbol{A}\) of \(6 \times 37^2 = 8214\) anchors. 

For each anchor \(a\) of each \(x_i\), Layer 19 maps \(a\) to two vectors. \texttt{Anchor\_Cls\_Conv} transforms \(a\) into the \textit{objectness} score \(\vec{o_p} = \) [\texttt{bkg\_score}, \texttt{source\_score}]. \texttt{Anchor\_Reg\_Conv}  transforms \(a\) into the \textit{anchor-source offset} \(\vec{\boldsymbol{d}} =\) [\(d_1, d_2, d_3, d_4\)]. Given anchor \(a\)'s spatial extent (\(a_x, a_y, a_w, a_h\)), Eq. \ref{eq:as_offset}~\citep{girshick2014rich} takes \(\vec{\boldsymbol{d}}\) as input, and outputs the spatial extent --- centre coordinates, width, and height --- of the RoI proposal. Therefore, \(\vec{\boldsymbol{d}}\) essentially predicts how \(a\) ought to be shifted and scaled to become a RoI proposal --- surrounding some source inside its bounding box.
\begin{equation}
\label{eq:as_offset}
S(\vec{\boldsymbol{d}}\:{;}\: a) = \big(d_1a_w + a_x,\text{  }d_2a_h + a_y,\text{  }e^{d_3}a_w,\text{  }e^{d_4}a_h\big)
\end{equation}

Both transformations in Layer 19 can be expressed by a fully connected layer, performing dot products between its weight vector \(\vec{w_j}\) and \(x_i\), where \(j = 1 ... 6m\), and \(|\vec{w_j}|=512\). We let \(m = 2\) for \texttt{Anchor\_Cls\_Conv} and \(m = 4\) for \texttt{Anchor\_Reg\_Conv}. In practice, these two transformations are implemented using \(6m\) filters of \(1 \times 1 \times 512\) convolutions for improved performance and efficiency. This is shown in Layer 19 (\texttt{Anchor\_Cls\_Conv} and \texttt{Anchor\_Reg\_Conv}) in Table \ref{tab:network_architecture}.




To train \texttt{Anchor\_Cls\_Conv} and \texttt{Anchor\_Reg\_Conv}, the RPN relies on \texttt{Anchor\_Target} to dynamically generate ground truths for each anchor \(a \in \boldsymbol{A}\). The ground truth for the \textit{objectness} score vector is a scalar \(o_{g}\), denoting a negative anchor by 0 or a positive by 1. It indicates whether \(a\) matches a nearby ground-truth box (generated in Section \ref{sec:train_test_set}) \(b\), and its quantity is determined by the Intersection-over-Union (IoU) overlap \(\frac{a \cap b}{a \cup b}\). \(a\) is positive if either (1) it has an IoU higher than a threshold \(\tau\) with any ground-truth boxes or (2) it has the highest IoU if no anchors are positive. We set \(\tau\) to 0.7 as a reasonable balance between loose (e.g. 0.5) and tight (e.g. 0.9) overlap values. An anchor is negative if its highest IoU overlap (with some ground-truth box) is less than \(1-\tau\), i.e. 0.3 in our tests. Anchors that are neither positive nor negative are excluded from training. Random selection is used to ensure the total number of negative and positive anchors is equal to the batch size \(B = 256\) for each subject. Moreover, efforts were made to keep the ratio between the positive and the negative roughly at \(1:1\) to avoid unbalanced training sets. The loss function for training \texttt{Anchor\_Cls\_Conv} against each batch is defined as:
\begin{equation}
\label{eq:loss_anch_cls}
 L_{ac} = \frac{1}{B}\sum_{i=1}^{B}\Big\{-\big[\log{\big(\texttt{softmax}(\vec{o_{p_i}})\big)}\bigcdot\texttt{one\_hot}(o_{g_i})\big]\Big\}
\end{equation}
where function \texttt{softmax}\((\cdot)\) converts \(\vec{o_{p_i}}\) into a  probability distribution, and function \texttt{one\_hot}\((\cdot)\) encodes the scalar \(o_{g_i}\) into a vector.

The ground truth for the predicted anchor-source offset vector \(\vec{\boldsymbol{d}}\) is calculated using the inverse of \(S\) defined as:
\begin{equation}
\label{eq:ap_gt}
\begin{aligned}
S^{-1}(\vec{\boldsymbol{b}}\:{;}\: a) &= \big(\frac{b_x - a_x}{a_w},\: \frac{b_y - a_y}{a_h},\: \log{\frac{b_w}{a_w}},\: \log{\frac{b_h}{a_h}}\big) \\
&= (g_1, g_2, g_3, g_4) = \vec{\boldsymbol{g}}
\end{aligned}
\end{equation}
Given anchor \(a\) and its spatial extent (\(a_x, a_y, a_w, a_h\)), Eq. \ref{eq:ap_gt} takes as input the spatial extent vector \(\vec{\boldsymbol{b}}\) of a ground-truth box \(b\), with which \(a\) has the highest IoU among all ground-truth boxes, and outputs the true (actual) anchor-source offset \(\vec{\boldsymbol{g}} = [g_1, g_2, g_3, g_4]\). The loss function for training \texttt{Anchor\_Reg\_Conv} is defined as:
\begin{equation}
\label{eq:loss_anch_reg}
L_{ar} = \frac{1}{|A'|}\sum_{i=1}^{|A'|}\Big(\sum_{j=1}^{4}\big[\texttt{smooth\_L1}(d_j - g_j)\:o_{g_i}\big]\Big)
\end{equation}
in which \(A' \subset \boldsymbol{A}\), and \(|A'| = 5241\). \(\boldsymbol{A}\setminus A'\) includes anchors that lie (partially) outside the subject, and function \texttt{smooth\_L1} is a Huber loss ~\citep{huber1964robust}.

\subsubsection{RoI Proposal}
\label{sec:roi_proposal}
In the second step of \texttt{LocNet}, the \texttt{RoI\_Proposal} layer shifts every anchor \(a \in \boldsymbol{A}\) by \(\vec{\boldsymbol{d}}\) based on Eq. \ref{eq:as_offset}, yielding \(6\times37^2\) candidate RoI proposals. After excluding unreasonably small candidates (i.e. less than \(4 \times 4\) pixels in the subject), it sorts remaining proposals by their \textit{source objectness} scores \texttt{softmax}\((\vec{o_p})[1]\) in a descending order, and selects the top \(M\) proposals (\(M\) is a hyper-parameter set to 6000) for pruning using the Non-Maximum Suppression (NMS) algorithm~\citep{neubeck2006efficient}. Iterating over the sorted list of \(M\) proposals, NMS accepts a proposal \(p'\) with the highest \textit{source objectness} score, then discards all subsequent proposals whose IoU overlap with \(p'\) is greater than a threshold (a hyper-parameter set to 0.7) and repeats the procedure with the remaining proposals until the end of the list. Finally only the top \(P\) scoring proposals are kept after NMS pruning, where \(P\) is a hyper-parameter set to 2000 and 5 for training and testing respectively.

During testing, each one of the 5 proposals \(p \in P\) is directly fed to the Recognition Network (\texttt{RecNet}) (cf. Section \ref{sec:rec_net}) to predict (1) the proposal-source offset \(\vec{\boldsymbol{u}}\), by which \(p\) ought to be shifted and scaled in order to become a nearby ground-truth box, and (2) the morphology class \(m\) (cf. Table \ref{tab:dataset_dist}) of \(p\). However, to train \texttt{RecNet} to perform such prediction during training, each one of the 2000 \(p \in P\) goes through the \texttt{RoI\_Proposal\_Target} layer, which aims to produce ground truths for both \(\vec{\boldsymbol{u}}\) and \(m\). For each \(p \in P\) and given a set \(T\) of ground-truth boxes associated with the subject, the ground-truth box \(t \in T\) that has the highest IoU with \(p\) is the \textit{target} of \(p\). The ground-truth of \(\vec{\boldsymbol{u}}\) for \(p\) is then calculated as:
\begin{equation}
\label{eq:ps_gt}
\begin{aligned}
S^{-1}(\vec{\boldsymbol{t}}\:{;}\: p) &= \big(\frac{t_x - p_x}{p_w},\: \frac{t_y - p_y}{p_h},\: \log{\frac{t_w}{p_w}},\: \log{\frac{t_h}{p_h}}\big) \\
&= (q_1, q_2, q_3, q_4) = \vec{\boldsymbol{q}}
\end{aligned}
\end{equation}
The ground truth of \(m\) is a scalar \(v \in \{0 ... 6\}\) denoting six morphology classes (\(1 - 6\)) plus the background class (0). However, since each \(t \in T\) contains a radio source with a given morphology defined in Table \ref{tab:dataset_dist}, \(v\) cannot possibly take the value of 0 to represent the background \textit{target}. To address this, the Faster R-CNN model treats as \textit{background} proposals the set \(G \subset P\) of proposals whose IoUs with their targets are within the range of [0.1, 0.5], and the ground truth of \(m\) for each \(g \in G\) is manually set to 0. Similarly, a proposal is \textit{foreground} if its IoU with its \textit{target} is greater than 0.5. Random selection is used to (1) adjust the number of \textit{foreground} and \textit{background} proposals such that the ratio between the two is approximately \(1:3\), and (2) to further reduce the total number of RoI proposals from 2000 to 128, thus \(|P| = 128\). The output of the \texttt{RoI\_Proposal\_Target} layer --- \(P\), and the ground-truth \(\vec{\boldsymbol{q}}\) and \(v\) associated with each \(p \in P\) --- is fed to \texttt{RecNet} for training.



\subsection{Recognition network}
\label{sec:rec_net}
For each subject, \texttt{RecNet} accepts two inputs --- (1) the feature map \(F\) produced by the convolution network layer \texttt{Conv5\_3} and (2) the set of RoI proposals \(P\) produced by either the \texttt{RoI\_Proposal} layer during testing or the \texttt{RoI\_Proposal\_Target} layer during training. For each \(p \in P\), the first layer of \texttt{RecNet} --- \texttt{ST\_RoI\_Pool} --- crops the RoI \(r\) out of \(F\) based on \(p\), and down-samples \(r\) into a feature map \(f\) of size \(512 \times 7 \times 7\).  The original Faster R-CNN \citep{ren2017faster} study uses RoI pooling~\citep{girshick2015fast} for down sampling. It works by evenly partitioning each channel of \(r\) into a \(7 \times 7\) grid of sub-sections, each of which has an approximate size \(37 / 7 \times 37 / 7\), and max-pooling the values from each sub-window to form a single channel of \(f\). However, the issue with RoI pooling is that while it accepts both \(F\) and \(P\) as input during forward pass, only the gradient w.r.t \(F\) is calculated during backpropagation via max pooling. The gradients w.r.t. \(P\) are completely ignored. In other words, training errors caused by \(P\) are not sufficiently accounted for, resulting in an approximate optimization solution at most. To overcome this limitation, we use two tensor operations defined in the Spatial Transformer Network~\citep{jaderberg2015spatial} to crop and down-sample \(r\) --- the affine transformation \( \mathcal{T}_\theta \) and the bilinear sampling \( \mathcal{B}\). Since \( \mathcal{T}_\theta \) is differentiable w.r.t. \(P\), and \( \mathcal{B}\) is differentiable w.r.t. both \(F\) and the output of \( \mathcal{T}_\theta \), the error gradients are able to flow back not only to \(F\) but also to coordinates of each \(p \in P\). Given the coordinates \([x_1, x_2, y_1, y_2]\) of \(p \in P\), the affine transformation is defined as:
\begin{equation}
\label{eq:affine_t}
\mathcal{T}_\theta(G_i) =
\begin{bmatrix}
    \frac{x_2 - x_1}{w_F}       & 0 & \frac{x_1 + x_2 - w_F}{w_F} \\
    \\
    0       & \frac{y_2 - y_1}{h_F} & \frac{y_1 + y_2 - h_F}{h_F}  \\
\end{bmatrix}
\begin{pmatrix}
    u^f_i\\
    v^f_i\\
    1
\end{pmatrix} = 
\begin{pmatrix}
    u^F_i\\
    v^F_i
\end{pmatrix} 
\end{equation}
where \(w_F = 37\) and \(h_F = 37\) are the width and height of \(F\), and \(G_i = (u^f_i, v^f_i)\ \forall i \in \{0,1,...,7^{2} - 1\}\) are coordinates of the regular grid on \(f\), and \((u^F_i, v^F_i) \) are coordinates of the sample points on \(F\). 

The output from \texttt{ST\_RoI\_Pool} is a set \(\boldsymbol{R}\) of RoI feature maps of size \(512 \times 7 \times 7\) and \(|M| = 128\) and \(5\) for training and testing respectively. The next fully-connected layer \texttt{FC\_6} reshapes \(\boldsymbol{R}\) as a matrix \(\boldsymbol{R'}\) of size \(|M| \times 25,088\), and uses a weight matrix of size \(25,088~\times~4,096\) to linearly transform \(\boldsymbol{R'}\) into a \(|M|~\times~4,096\) matrix \(F_1\). During training, a dropout layer~\citep{srivastava2014dropout} \texttt{Dropout\_6} is added such that for a given element \(el\) of \(F_1\), \texttt{Dropout\_6} either resets the value of \(el\) to 0 with a probability of \(1 - k\) or scales up the value of \(el\) by a factor of \(\frac{1}{k}\) with a probability of \(k, 0\leq k \leq 1\). Compared to conventional regularization methods, Dropout is more effective and computationally efficient to prevent overfitting for layers with a large number of parameters --- 102 million weights in the case of \texttt{FC\_6}. After dropout updates, \(F_1\) is transformed by another fully-connected layer \texttt{FC\_7} followed by another dropout layer \texttt{Dropout\_6}, producing a matrix \(F_2\) of size \(|M| \times 4,096\). It should be noted that dropout layers --- \texttt{Dropout\_6} and \texttt{Dropout\_7} --- are only used during training, and are skipped during testing as shown in Figure \ref{fig:loc_rec_net}. Both \texttt{FC\_6} and \texttt{FC\_7} use RELU as their internal activation function to output \(F_1\) and \(F_2\).

The first output of \texttt{RecNet} contains scores of each RoI \(r \in \boldsymbol{R}\) against morphology classes defined in the first row of Table \ref{tab:dataset_dist}. To produce such output, a fully-connected layer \texttt{FC\_Cls\_Score} takes \(F_2\) as input, and produces as output an \(|M| \times 7\) matrix \(F_3\), whose value at row \(i\) and column \(j\) denotes the score of the \(i\)th RoI in \(\boldsymbol{R}\) being an instance of class \(j\), and \(1\leq i < |M|,0\leq j \leq6\). During training, \(F_3\) is used as the input of the classification log-loss function \texttt{RoI\_Cls\_Loss} shown as the gray-rectangle at the bottom of Figure \ref{fig:loc_rec_net}. The formal expression of \texttt{RoI\_Cls\_Loss} \(L_{rc}\) is defined as:
\begin{equation}
\label{eq:loss_roi_cls}
 L_{rc} = \frac{1}{|M|}\sum_{i=1}^{|M|}\Big\{-\big[\log{\big(\texttt{softmax}(F_3[i])\big)}\bigcdot\texttt{one\_hot}(v_i)\big]\Big\}
\end{equation}
where scalar \(v_i \in \{0 ... 6\}\) denotes the ground truth class for the \(i\)th RoI in \(\boldsymbol{R}\), and is provided by the \texttt{RoI\_Proposal\_Target} layer as described in Section \ref{sec:roi_proposal}.
The \texttt{softmax} function in the \texttt{Cls\_SoftMax} layer converts the \(i\)th row of \(F_3\) into a discrete probability distribution vector \(\vec{d}\), whose \(j\)th element represents the probability of RoI \(i\) being an instance of class \(j\). In practice, the morphology class \(\hat{m}\) with the highest probability is often chosen as the output classification result.

The second output of \texttt{RecNet} contains the proposal-source offsets of each \(r \in \boldsymbol{R}\) for each morphology class. To produce such output, the \texttt{FC\_Reg\_Pred} layer takes \(F_2\) as input, and produces as output an \(|M| \times 28\) matrix \(F_4\), whose values at row \(i\)th and between columns \([4j, 4j + 4]\) denote the proposal-source offsets of the \(i\)th RoI for class \(j\), and \(1\leq i < |M|,0\leq j \leq6\). During training, \(F_4\) is used as the input of the regression loss function \texttt{RoI\_Reg\_Loss} (the rectangle at the bottom right of Figure \ref{fig:loc_rec_net}), which is defined as:
\begin{equation}
\label{eq:loss_roi_reg}
L_{rr} = \frac{1}{|M|}\sum_{i=1}^{|M|}\Big(\big[\texttt{smooth\_L1}(\vec{d_{ij}} - \vec{g_{ij}})\big]\Big)
\end{equation}
where \(\vec{d_{ij}} = [d_x, d_y, d_w, d_h]\) is the predicted proposal-source offset of RoI \(i\) corresponding to its true morphology class \(j\), \(\vec{g_{ij}} = [g_x, g_y, g_w, g_h]\) is the ground-truth proposal-source offset of \(i\) for the same true class \(j\), and \texttt{smooth\_L1} is the Huber loss function~\citep{huber1964robust}. 





\section{Quantifying classification precision}
\label{sec:evaluation}
We implement the data pipeline described in Section \ref{sec:data_pipeline} using Tensorflow~\citep{abadi2016tensorflow}. Both training and testing require GPU resources, and we deploy the pipeline to run on both Tesla K40c (12GB device RAM) and Tesla P100 (16GB device RAM) GPUs. For training, we use the momentum optimizer to update network weights, and set the initial learning rate to 0.001 with a decay rate of 0.1 for every 50,000 iterations. The training speed is about 0.52 seconds and 0.11 seconds per iteration on K40 and P100 respectively. Thus a pipeline instructed to execute 80,000 iterations requires 3 to 12 hours of training time on provisioned GPU resources. For testing, it takes the learned model 220 milliseconds and 45 milliseconds per subject on K40c and P100 respectively to generate detected radio sources, their associated morphology and probabilities.
\subsection{Training error}
\begin{figure}
	\includegraphics[width=\columnwidth]{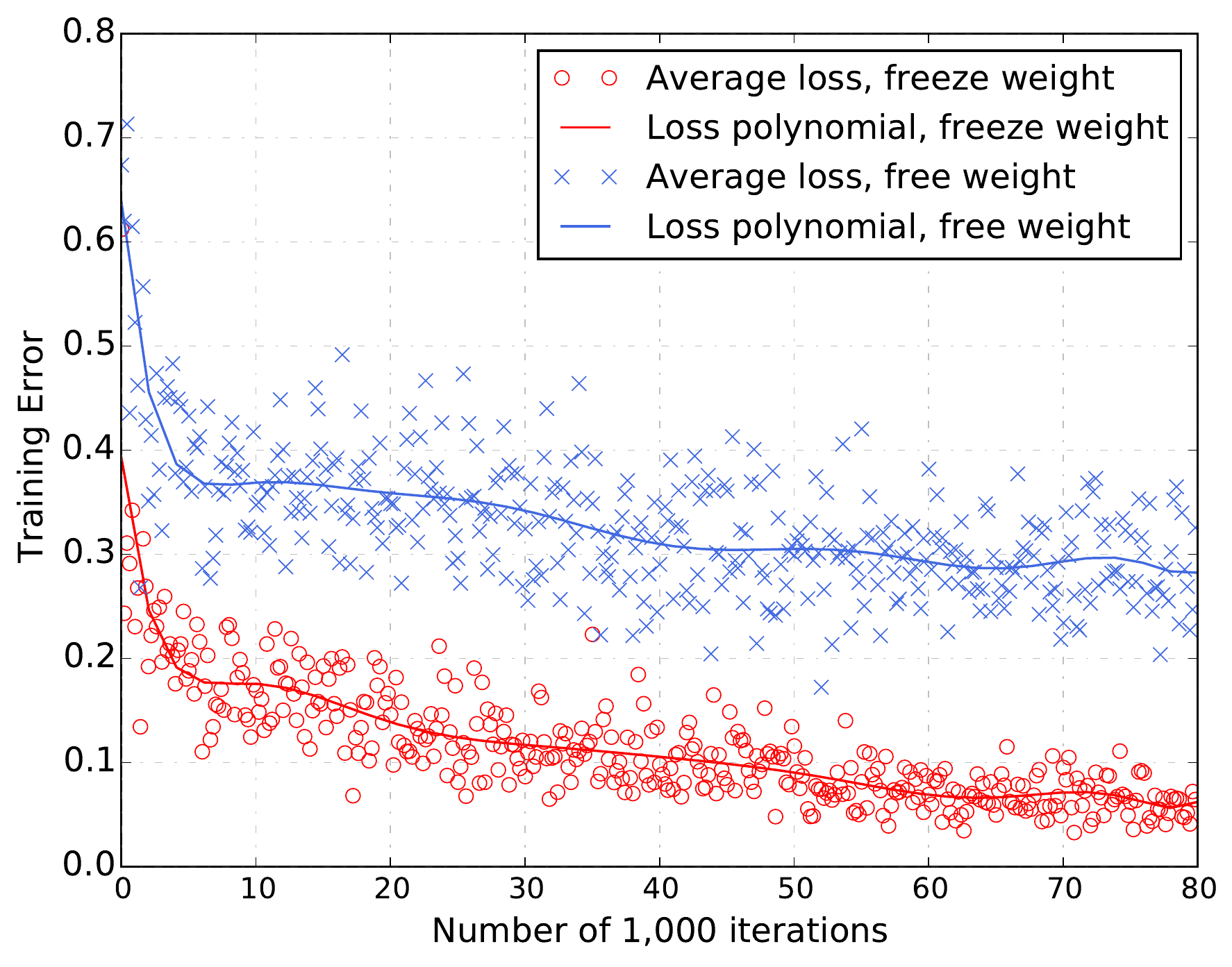}
    \caption{Learning curves (on dataset D4) monitor the change of training losses (Y-axis) as the training progresses by some number of iterations (X-axis). The top part denotes the case where the low-level (i.e. layer 1, 2, 4, and 5 in Table \ref{tab:network_architecture}) kernel weights (\textit{N}=259,716) are trainable --- i.e. free to be updated during the training process. The bottom part shows when those low-level kernels weights freeze and are thus not updated during training. The losses are sampled every 10 iterations during training, and are only shown every 200 iterations for visualization. However, both polynomials are plotted based on all collected loss samples.}
    \label{fig:learning_curve}
\end{figure}

The efficiency and effectiveness of the training pipeline is largely determined by the \emph{training error}, which is the sum of the four losses defined in Eq. \ref{eq:loss_anch_cls}, \ref{eq:loss_anch_reg}, \ref{eq:loss_roi_cls}, and \ref{eq:loss_roi_reg}:
\begin{equation}
\label{eq:training_error}
\text{Training Error} = L_{ac} + L_{ar} + L_{rc} + L_{rr}
\end{equation}
The goal of training is to reduce the training error on the training set using various optimization techniques 
without compromising the model generality on future unseen datasets. To examine the change of training error, we compare two learning curves in Figure \ref{fig:learning_curve}, where the Y-axis denotes training errors and the X-axis represents the number of iterations. As training proceeds on dataset D4, the average training error becomes smaller in both cases, reduced from 0.35 to 0.05 for the bottom learning curve, and from 0.7 to 0.28 for the top curve. Both curves exhibit a sharp plunge within the first 5,000 iterations, and turn into a more steady descent afterwards. The downwards trend appears to plateau out after 65,000 to 75,000 iterations for both curves, suggesting the model has reached its learning capacity given current network architecture and datasets.

Training errors in the bottom learning curve in Figure \ref{fig:learning_curve} are significantly smaller than those in the top curve. 
The bottom learning curve was generated by the training process in which low-level (i.e. layer 1, 2, 4, and 5 in Table \ref{tab:network_architecture}) convolutional kernels were set to read-only once loaded from the pre-trained VGG-16 model, and were never updated throughout training. The training process that produced the top curve, on the other hand, continuously updates these low-level kernels during training. Since these low-level kernels have been pre-trained using much larger datasets for an extended period of time (e.g. several weeks), we believe they capture features common enough to be shared across different domains. 

Figure \ref{fig:learning_curve} suggests that freezing these low-level kernels in effect reduces the training error with a much higher efficiency. This is because pre-trained low-level weights become fine-tuned and optimal in extracting low-level features \textit{common} to generic object detection tasks including those in {\cla}. If not retained during re-training (particularly given the high initial learning rate and different loss functions), they are subject to gradient updates much higher than those received towards the end of the ImageNet pre-training. Consequently, they quickly diverge from the current optimal region in the high-dimensional parameter space.
\begin{figure}
	\includegraphics[width=\columnwidth]{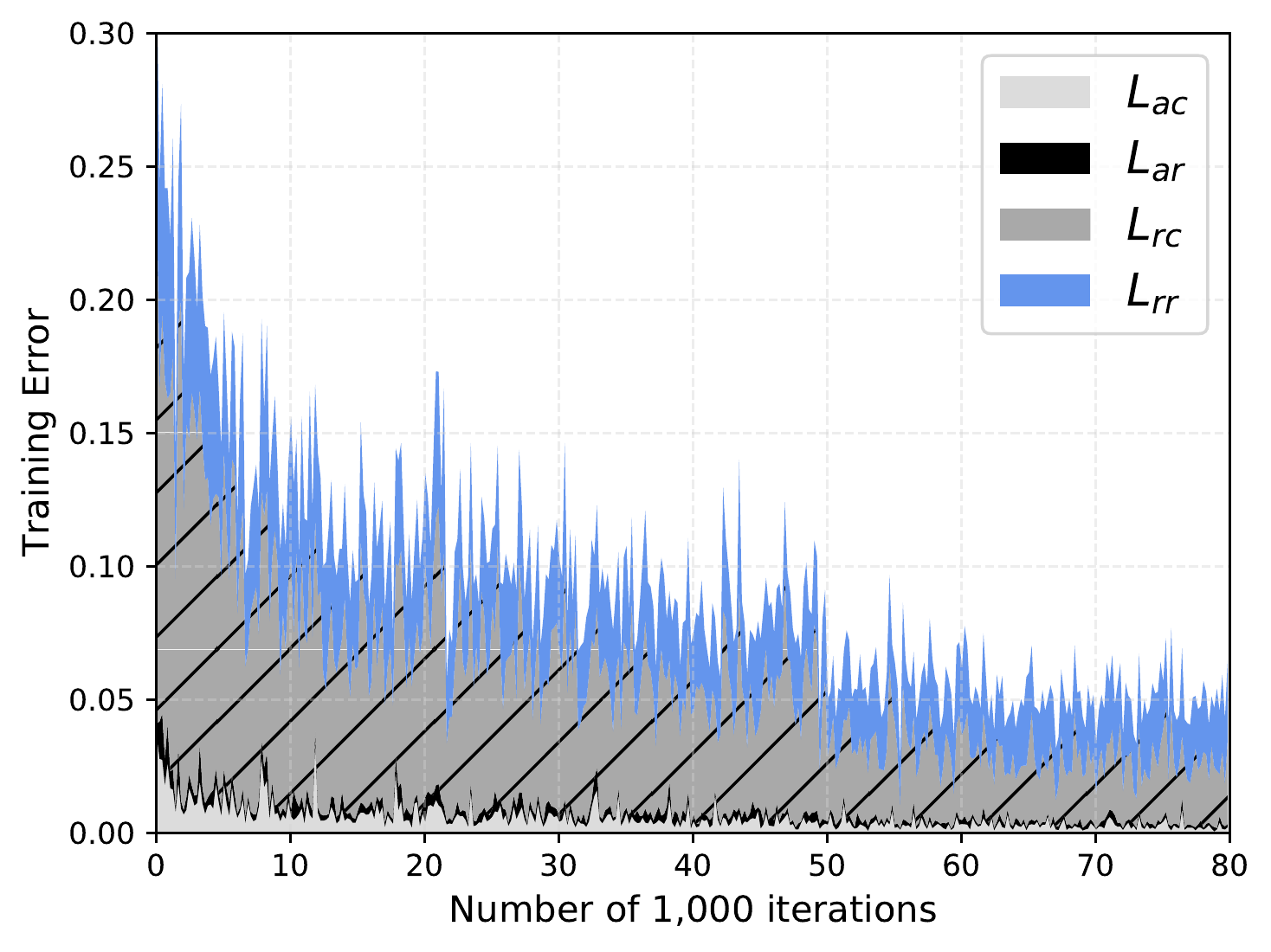}
    \caption{Training errors are decomposed into four losses~(Eq. \ref{eq:training_error}) stacked on top of one another every 10 iterations as the training progresses. Areas covered by dark diagonal lines denote the RoI classification loss $L_{rc}$.}
    \label{fig:error_decomp}
\end{figure}

Since the training error defined in Eq. \ref{eq:training_error} is the sum of four loss terms, we visually break down the training error as a stack plot shown in Figure \ref{fig:error_decomp}. Initially, about 60\% of the training error was attributed to the RoI classification loss \(L_{rc}\). While the overall training error declines as training progresses, the portion of \(L_{rc}\) is gradually diminishing, reaching to 35\% in the end. On the other hand, the portion of \(L_{rr}\) is increasing to above 55\%. 
This suggests that training of morphology classification is slightly more efficient than that of localization regression. We find that the correlation coefficients
between Anchor errors (\(L_{ac}\) and \(L_{ar}\)) and \(L_{rr}\) are slightly higher than those between Anchor errors (\(L_{ac}\) and \(L_{ar}\)) and \(L_{rc}\), suggesting RoI regression is more sensitive to errors caused by the region proposal network. 
Moreover, there is a moderate positive correlation (0.508) between \(L_{rc}\) and \(L_{rr}\) since these two tasks share a large number of weights in the fully-connected layers 24 and 26, which contain 87.4\% of the parameters stored in the model.

\subsection{Testing metrics and evaluation}
\label{sec:metrics_eval}
To evaluate {\cla} against the testing set, we use a single evaluation metric --- the mean Average Precision (mAP). The Average Precision (AP) is a function of both {\em{reliability}} and {\em{completeness}}, which are referred to as {\em{precision}} and {\em{recall}} respectively in machine learning. 
{\em{Precision}} measures the fraction of identified sources that are correct according to the RGZ ground truth and {\em{Recall}} refers to the fraction of RGZ ground-truth radio sources that have been identified.
Given a morphology class \(m \in {1...6}\), let \(L_m\) denote a list of radio sources detected by {\cla} as ``class \(m\) sources" from all subjects in the testing set, and let \(T_m\) denote a set of radio sources that are truly of morphology \(m\) contained in the testing set. Sources in \(L_m\) are ranked by their morphology class probabilities ($P$-values) in a descending order. The Average Precision \(AP_m\) for morphology class \(m\) is calculated as:
\begin{equation}
\label{eq:average_precision}
AP_m = \frac{\sum_{k=1}^{|L_m|}(P(k)\times \text{tp}(k))}{|T_m|}
\end{equation}
where \(\text{tp}(k)\) is an indicator function equaling 1 if \(L_m[k]\) is a true positive detection, 0 otherwise, and \(P(k)\) denotes the {\em{precision}} calculated up to element \(L_m[k]\):
\begin{equation}
\label{eq:precision_at_k}
P(k) = \frac{\sum_{i=1}^k(\text{tp}(i))}{k},\quad 1\leq k \leq |L_m|
\end{equation}
A detected source \(K \in L_m\) in subject \(S\) is true positive if and only if the IoU (defined in Section \ref{sec:rpn}) between \(K\) and some ground-truth sources of class \(m\) in \(S\) is greater than 0.5.

Finally the mean Average Precision (mAP) is calculated as:
\begin{equation}
\label{eq:mean_average_precision}
mAP = \frac{\sum_{m=1}^6(AP_m)}{6}
\end{equation}
We apply Eq. \ref{eq:average_precision} and Eq. \ref{eq:mean_average_precision} to evaluate the testing set detection results produced by five different data pre-processing methods --- F, D1, D2, D3, and D4 as discussed in Figure \ref{fig:data_prep}. The result of both AP and mAP for each method is presented in Table \ref{tab:metrics_eval}.

\begin{table*}
\large
\begin{center}
\caption{Evaluation of 5 data pre-processing methods using AP and mAP. Each row represents APs achieved by all five methods for a given morphology class. The highest AP for each morphology class is highlighted in the bold face. Each column denotes APs achieved by a particular method over all six morphology classes and the overall mAP. Method D4 has achieved the highest mAP, highest APs for morphology 1C\_1P and 2C\_2P, and second highest AP for 3C\_3P.}
  \label{tab:metrics_eval}
  \begin{tabular}{r | c | c | c | c | c |  }
    \hline
    Methods & \(F\) & \(D1\) & \(D2\) & \(D3\) & \(D4\) \\ \hline
     \hline
      & 
         \includegraphics[scale=0.7]{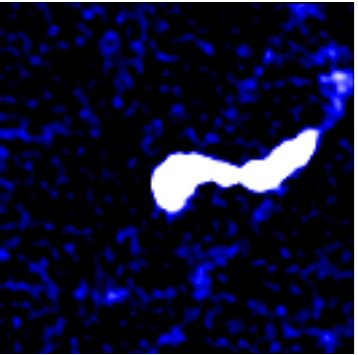} 
          & \includegraphics[scale=0.7]{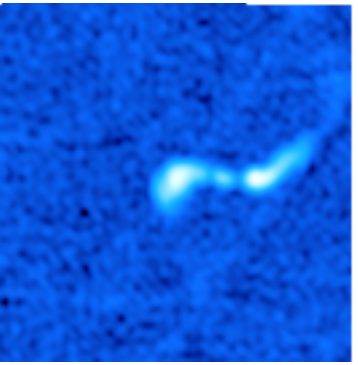} & \includegraphics[scale=0.7]{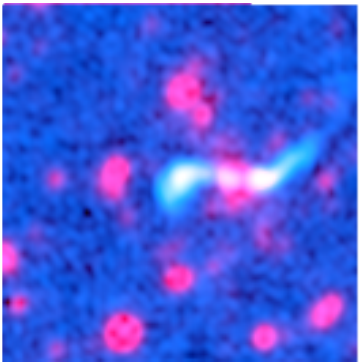} & \includegraphics[scale=0.51]{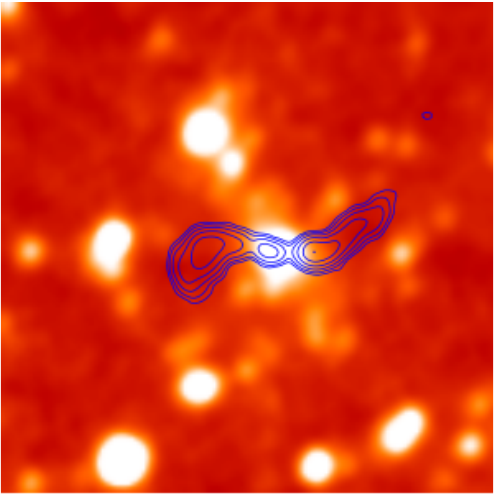} & \includegraphics[scale=0.51]{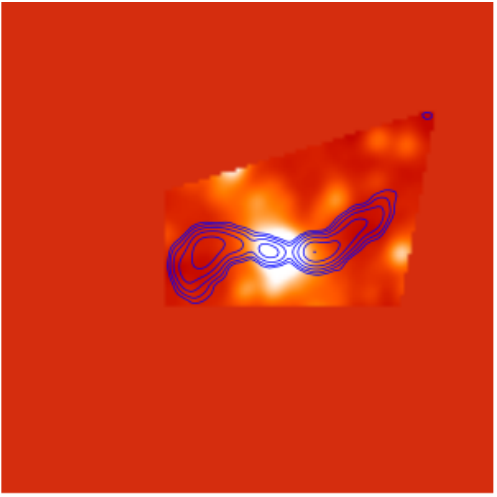}\\
     \hline
    1C\_1P & 0.8087 & 0.8580 & 0.8242 & 0.8485 & \textbf{0.8784}\\ 
    \hline
    1C\_2P & 0.6376 & 0.6882 & 0.6843 & 0.6746 & \textbf{0.7074}\\ 
    \hline
    1C\_3P & 0.8250 & 0.8816 & 0.8561 & 0.8876 & \textbf{0.8941}\\ 
    \hline
     2C\_2P & 0.7474 & 0.7014 & 0.7231 & 0.7983 & \textbf{0.8200}\\ 
    \hline
     2C\_3P & \textbf{0.8087} & 0.7099 & 0.6989 & 0.8047 & 0.7916\\ 
    \hline
    3C\_3P & 0.7708 & 0.8636 & 0.8561 & \textbf{0.9424} & 0.9269\\ 
    \hline
    mean AP & 78.5\% & 78.4\% & 77.4\% & 82.6\% & \textbf{83.6}\%\\ 
    \hline
    
     \end{tabular}
\end{center}
\end{table*}
The results of F and D1 --- pure radio emission --- are slightly better than D2, which simply places spatially-aligned radio and infrared planes in different channels of the input subject. This suggests that radio source detection from multi-wavelength datasets requires different data fusion techniques than those used for object detection from common RGB images. We therefore explore several alternative data fusion methods, and found methods D3 and D4 have consistently achieved better AP and mAP than other methods. On the other hand, not all fusion methods worked as expected. For example, in one method, we prepend to the network a \(1 \times 1 \times 3\) convolutional layer \cite{szegedyrabinovich2015}, which is then trained to learn optimal weighted averages of fluxes from different channels in the original subject input. However, this method is merely 0.5\% better than D2, achieving a mAP of 77.9\%. We suspect the reason D3 and D4 perform better is because their fusion method visually resembles the RGZ Web interface, through which citizen scientists have collectively produced the `RGZ truth' for training {\cla}. However, we note that visual classification may not always reflect the 
`true' ground truth as the accuracy of the classifications may be limited by the angular
resolution, frequency or sensitivity of the observations.  However, the purpose of our work
is to be able to replicate the accuracy standards set by visual classifications in an automated fashion. 

\subsection{Reliability versus Completeness}
\begin{figure}
	\includegraphics[scale=0.48]{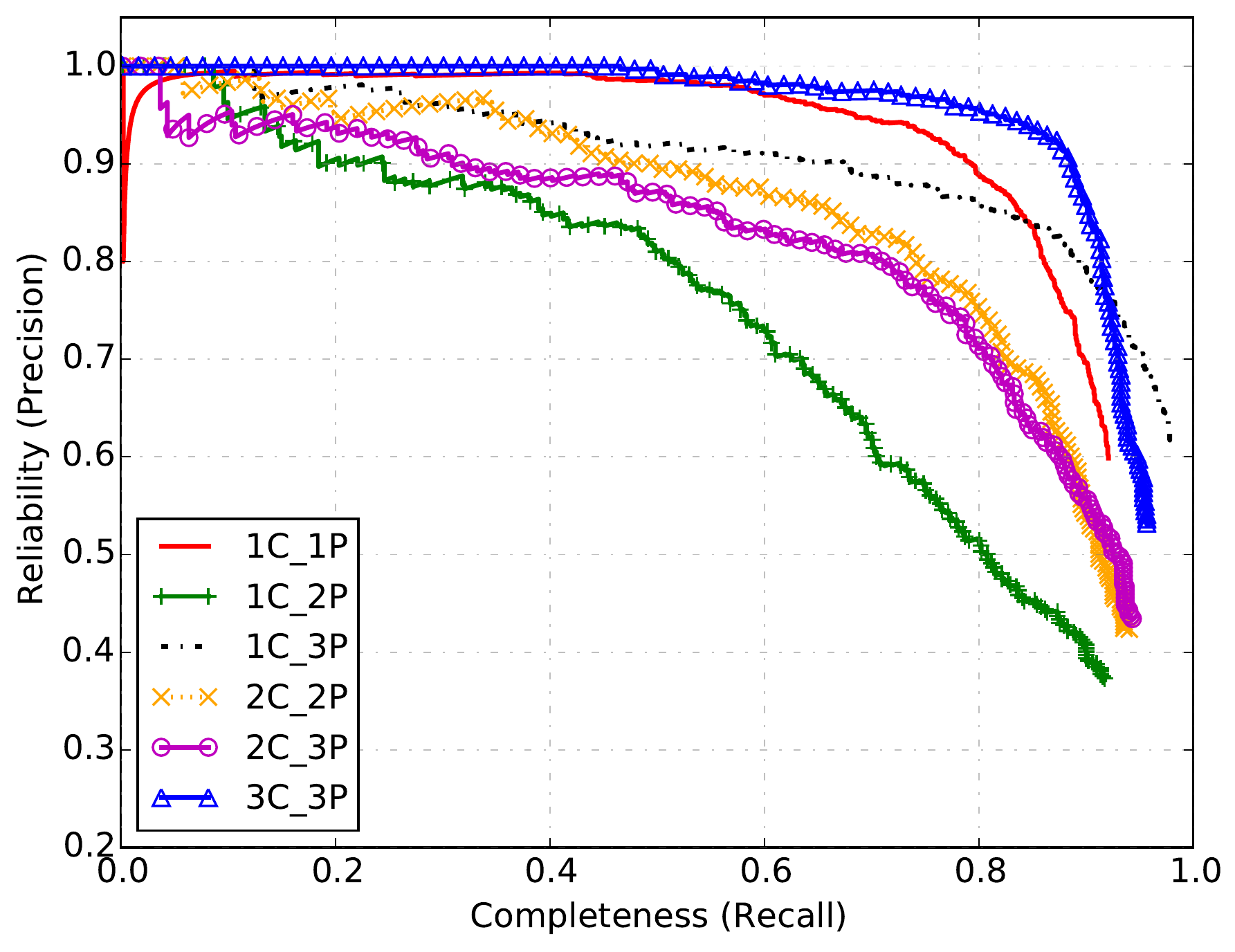}
    \caption{The trade-off between reliability and completeness is shown by the PR curves against the unseen test dataset with 4,603 subjects using the pre-processing method D4. Each morphology class has its own PR curve, which records the reliability (Y-axis) achieved by {\cla} (using method D4) at each level of completeness (X-axis). The area under a PR curve is known as Average Precision (AP), which has a discrete form expressed in Eq. \ref{eq:average_precision}. Therefore, the top right curves, with larger area beneath them, have greater APs.
    }
    \label{fig:prec_curve}
\end{figure}
The Precision-Recall (PR) curves plotted in Figure \ref{fig:prec_curve} shows how {\cla} deals with the trade-off between these two metrics for different morphology classes. In general, PR curves closer to the top right corner (e.g. 3C\_3P) have better mAPs than those further away from it. The 3C\_3P PR curve starts with a horizontal line (at the reliability level of 1.0) until the completeness level reaches 0.6. This suggests that, if we put all predicted 3C\_3P sources into a list $L$ sorted by $P$-values in descending order, and let $C$ be the set of ground-truth 3C\_3P sources in the testing set (where $|C|=668$ as per Table \ref{tab:dataset_dist}), then 60\% of $C$ ($N$ = 400) are also the \textit{first} 400 sources in $L$, and 80\% of $C$ ($N$ = 534) are in the \textit{first} 561 elements of $L$.

In contrast, in the PR curve for 1C\_2P, 
the reliability quickly drops immediately after 30\% of the true 1C\_2P sources have been detected, and by the time the completeness reaches 80\%, nearly half of the detected 1C\_2P sources are false positives. This is consistent with the relatively poor mAP results shown in Table \ref{tab:metrics_eval}. In particular, the wiggle section between Completeness 0.1 and 0.2 of the PR curve is caused by some top-ranked yet false positive 1C\_2P detections. In general, false positives have lower $P$-values because most PR curves in Figure \ref{fig:prec_curve} are smoothly bent downwards to the right. 
\begin{figure}
	\includegraphics[width=\columnwidth]{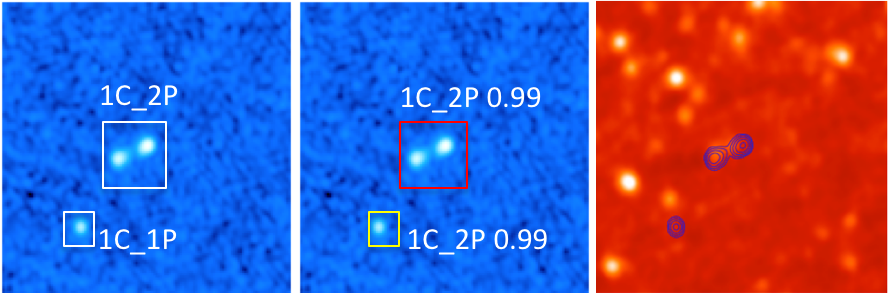}
    \caption{A `mis-classified' source 1C\_2P (bottom left) in subject \texttt{FIRST J131100.4+034608} selected from the training set. From left to right are: RGZ truth (white boxes), Sources detected by {\cla} (colored boxes), and \(5\sigma\) radio contours overlaid on the IR map. The bottom left source has a high probability (99\%) of being 1C\_2P, which should have been 1C\_1P according to the RGZ truth. False positive detections such as this one (with a high $P$-value) will cause the sudden drop of the 1C\_2P PR curve shown in Figure \ref{fig:prec_curve}.}
    \label{fig:culprit01}
\end{figure}

To identify potential causes for this, we show several false positive 1C\_2P examples taken from the training set.
Figure \ref{fig:culprit01} shows {\cla} outputs for two sources: a true positive 1C\_2P with a high $P$-value of 99\% at the centre, and a false positive 1C\_2P at the lower left with an equally high probability. 
It appears that this source is slightly elongated, but 
\begin{figure}
	\includegraphics[width=\columnwidth]{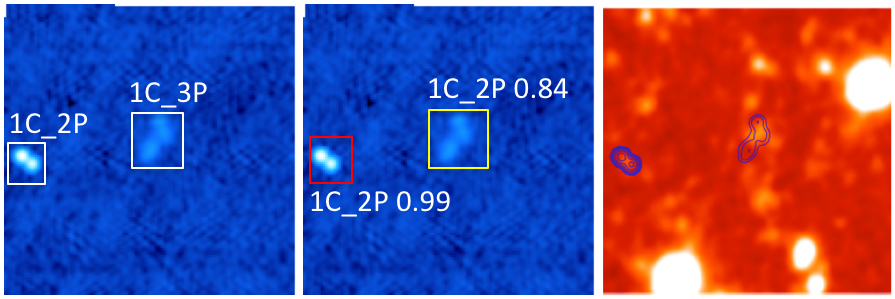}
    \caption{A `mis-classified' 1C\_2P source in the middle of subject \texttt{FIRST J110148.2+252746} with a relatively high probability of (84\%). According to the RGZ truth, it should have been 1C\_3P. But this mistake is more likely due to differences in the contour level between the RGZ DR1 pipeline and the {\cla} data preparation.}
    \label{fig:culprit02}
\end{figure}
it should be noted that `ground-truth peaks' did not come from RGZ user consensus but were automatically produced by the RGZ DR1 pipeline. The false detection in Figure \ref{fig:culprit02} could be caused by the difference in the contour level (\(4\sigma\)) used in DR1 and that  (\(5\sigma\)) used for training {\cla}. This difference may prevent {\cla} from distinguishing the two peaks at the top right. However, we find that laying \(4\sigma\) contours to train {\cla} exposes more unrelated noise in general, jeopardizing the overall detection performance. Our tests show that the D4 method could only achieve an mAP of 78\% when using \(4\sigma\) contours. 
These two examples show that resolving double peaks from a relatively small single-component source (1C\_2P) poses challenges to {\cla}, 
which could potentially confuse a star forming galaxy with an AGN. Identifying triple peaks from a double-component source (2C\_3P) also appears challenging to {\cla}.
\begin{table}
    \large
	\centering
	\caption{Evaluation of D3 and D4 based on a three-class scheme, in which only the ground-truth ``\# of components" is used to determine the classification of each radio source in the testing set.}
	\label{tab:component_map}
	\begin{tabular}{rcc} 
		\hline
		Morphology class & \(D3\) & \(D4\) \\
		\hline \hline
	    1C\_[1P or 2P or 3P] & 0.8644 & 0.9054 \\
		2C\_[2P or 3P] & 0.8699 & 0.8946 \\
	    3C\_3P & 0.9424 & 0.9269 \\
		\hline
		 mAP & 89.2\% & 90.9\% \\
	\end{tabular}
\end{table}

Although {\cla} does not agree with the RGZ truth in terms of the number of peaks for certain 1C\_2P and 2C\_3P sources, we hypothesize that {\cla} is able to correctly identify their components as exemplified in Figures \ref{fig:culprit01} and \ref{fig:culprit02}. 
To verify this hypothesis, we re-organize sources in the testing set into three morphology classes based on their ground-truth ``\# of components" regardless of their "\# of peaks". We then re-categorize sources detected by {\cla} from 6 classes (as in the first column of Table \ref{tab:metrics_eval}) into 3 classes based solely on their ``\# of components". For example, sources of classes 1C\_1P, 1C\_2P and 1C\_3P are merged into a single class denoted by 1C\_[1P or 2P or 3P]. Finally, we use Eq. \ref{eq:average_precision} and Eq. \ref{eq:mean_average_precision} to evaluate \(D3\) and \(D4\) against these three classes instead of the original six classes. The result is shown in Table \ref{tab:component_map}. All metrics in Table \ref{tab:component_map} are higher than those in Table \ref{tab:metrics_eval} (except for 3C\_3P that remains unchanged), particularly for 1C\_2P and 2C\_3P. This indicates that {\cla} is able to produce correct components for most of the 1C\_2P and 2C\_3P sources, increasing overall mAPs by nearly 8\% for both D3 and D4. In practice, we can recover ground-truth peaks by re-running the same peak calculation algorithm used in DR1 on each RoI detected by {\cla}. Since this paper focuses on the development and evaluation of a deep learning method, we leave for future work the optimal integration of {\cla} with other RGZ data reduction and analysis algorithms.

\subsection{Multi-source subjects}
A key problem that RGZ aims to address is to distinguish multiple unrelated sources from multiple components of single sources. {\cla} demonstrates this capability in Figure \ref{fig:detection_examples}$A$, Figure \ref{fig:culprit01} and \ref{fig:culprit02} (regardless of peaks). However, a statistical measure is needed to quantify this capability. Since 94\% of the subjects in the testing set (4,858 sources in 4,603 subjects) have only one radio source, mAPs in Table \ref{tab:metrics_eval} and \ref{tab:component_map} do not effectively measure {\cla}'s performance in separating multiple sources. Therefore, we create a `small' data set $T$, which includes \textit{every} subject in the testing set that has at least 2 sources. In total, $T$ contains 505 such sources, excluding 4,353 single-source subjects (i.e. 4,353 sources) from the original testing set.
\begin{table}
    \large
	\centering
	\caption{Evaluation of D3 and D4 in a `small' (250 subjects) testing set $T$, in which each subject has at least 2 RGZ DR1 sources within its 3-arcmin by 3-arcmin FoV. The first column denotes the number of sources with the corresponding number of components.}
	\label{tab:eval_multisource}
	\begin{tabular}{rrcc} 
		\hline
		Source count & Morphology & \(D3\) & \(D4\) \\
		\hline \hline
	    487 & 1C\_ & 0.7452 & 0.8394 \\
		13 & 2C\_ & 0.2800 & 0.3892 \\
	    5 & 3C\_ & 0.8850 & 0.2709 \\
		\hline
		505 & mAP & 63.7\% & 50.0\% \\
	\end{tabular}
\end{table}

Table \ref{tab:eval_multisource} presents mAPs that are significantly lower than those in Table \ref{tab:component_map}. Although D3 achieves a reasonably good AP (0.88) on 3-component (3C) sources, it performs very poor on 2-component (2C) sources (0.28). While D4 has a marginally improved 2C AP (0.38), its 3C AP is low. This shows that identification of multi-component sources from multi-source subjects still poses a challenge to {\cla}. However, it is worth noting that the median CL of 2C and 3C sources in $T$ is merely 0.64. 
Moreover, given the low number of sources (18) of classes 2C and 3C in $T$, their APs do not constitute reliable statistical measures, and this is particularly true for 3C. Given that the RGZ DR1 (with more than 11,000 multi-component sources with CL > 0.6) has the largest set of multi-component radio sources that have been visually classified and labelled to date, we need to obtain additional datasets with far more multiple-source subjects to obtain quantitative measures. This will be the main focus for our future work, which will update Table \ref{tab:eval_multisource} based on a larger number of multi-source subjects in the testing set. 


\subsection{Predicted box sizes}
\begin{figure}  
	\includegraphics[width=\columnwidth]{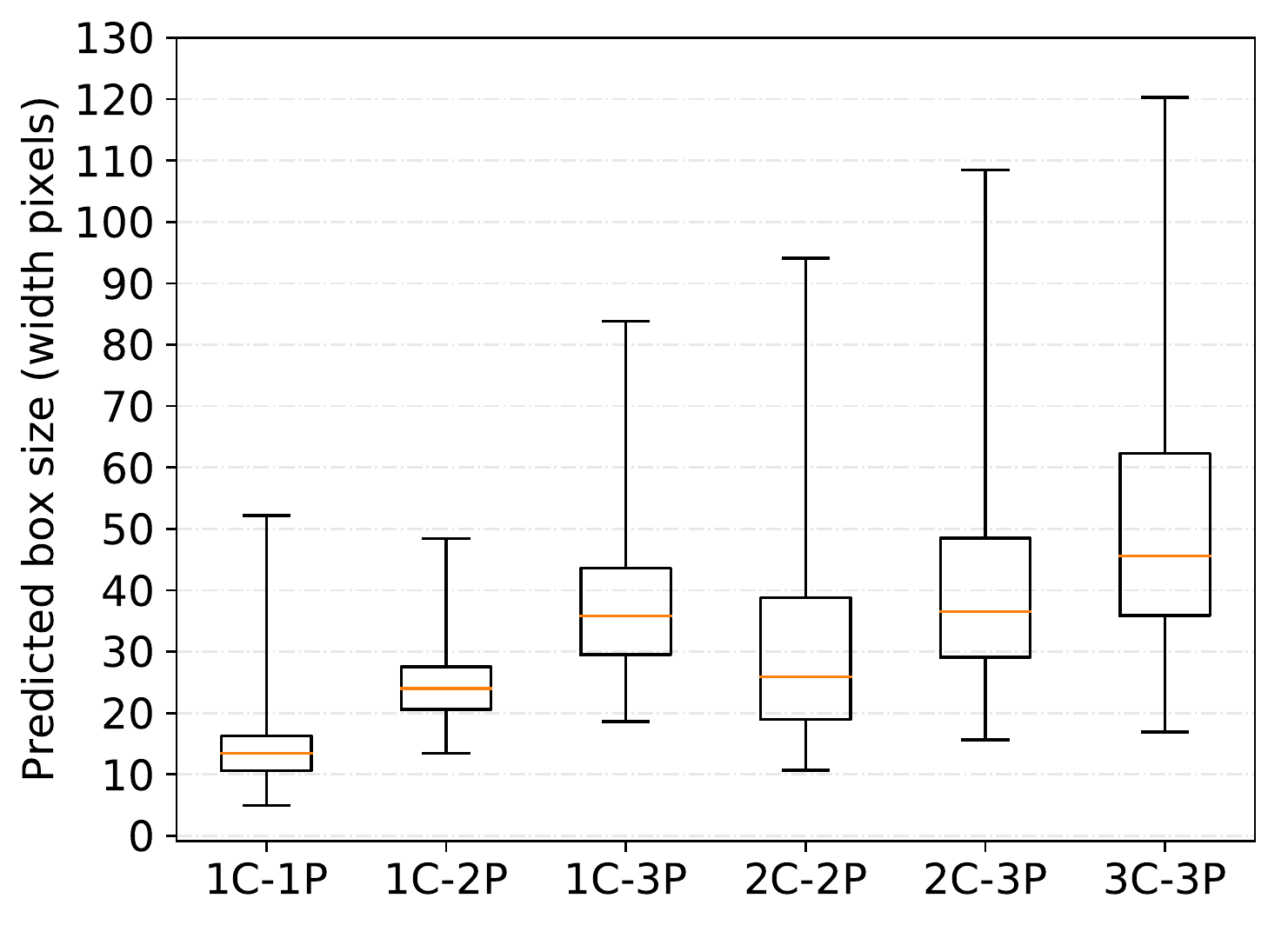}
    \caption{The distribution of detected box sizes for each class in the testing dataset. Similar to Figure \ref{fig:box_size}, each box spans the third and the first quartile size for a morphology.}
    \label{fig:pred_box}
\end{figure}

Since the RoI regression loss contributes 55\% of the total training error as shown in Figure \ref{fig:error_decomp}, we compare the box sizes detected by {\cla} and the box sizes specified in the RGZ truth in the testing set. Figure \ref{fig:pred_box} shows the size distributions of detected boxes for each morphology class in the testing set. They appear visually consistent in terms of medians and interquartile ranges with Figure \ref{fig:box_size}. But how do they compare to the testing-set ground truth? 
We calculate the correlation coefficients between the size (width) of each {\cla}-generated box and the size of its matching ground-truth (DR1) box. Table \ref{tab:box_size_corr} shows that the correlation coefficients are high (> 0.97) across all 6 morphology classes for both D3 and D4. This suggests that box sizes predicted by {\cla} are very close to ground-truth values for all six morphologies.

\begin{table}
\large
\centering
\caption{Correlation coefficients between sizes of DR1 (ground-truth) bounding boxes and sizes of boxes predicted by \cla\ for subjects in the testing set.}
\label{tab:box_size_corr}
\begin{tabular}{lcc}
\hline
Morphology & $D3$ & $D4$\\
\hline \hline
1C\_1P & 0.9718 & 0.9712 \\
1C\_2P & 0.9877 & 0.9866\\
1C\_3P & 0.9933 & 0.9946\\
2C\_2P & 0.9940 & 0.9952\\
2C\_3P & 0.9934 & 0.9916\\
3C\_3P & 0.9939 & 0.9927\\
\hline
\end{tabular}
\end{table}

\begin{figure}
	\includegraphics[width=\columnwidth]{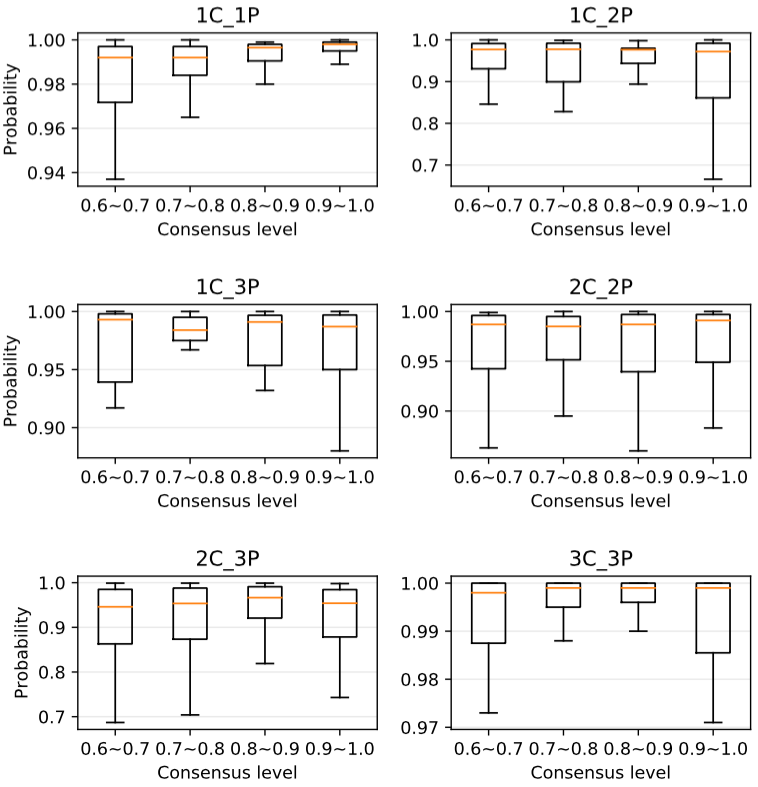}
    \caption{Comparisons between RGZ consensus levels (\(X\)-axis) and {\cla} classification probabilities (\(Y\)-axis) in the D4 testing set. The consensus level is segmented into four bins. Given a morphology class \(m\) within each bin, the box plot shows the distribution of classification probabilities of true positive detections whose morphology is \(m\).}
    \label{fig:prob_cl}
\end{figure}
\subsection{P-value versus Consensus level}
In order to ascertain whether RGZ consensus levels might have affected {\cla}'s performance, we examine the distribution of classification probabilities (\(P\)-values) of radio sources based on their RGZ consensus levels as shown in Figure \ref{fig:prob_cl}. Intuitively, a higher level of consensus corresponds to an easier case, which in turn should result in a more ``confident" classification result. This is indeed the case for simple morphology 1C\_1P, as CL increases from 0.6 to 1.0, the inter-quartile range (IQR) becomes much smaller, thus producing more stable and robust classifications, although the increase of median \(P\)-value is negligible: \(\leq 1\%\). However, the reduction of IQR is because 50\% of 1C\_1P sources have a CL close to 1 (as shown in Figure \ref{fig:cl_dist}) and the total number of 1C\_1P sources is substantially greater than other classes (as shown in Table \ref{tab:dataset_dist}). 

It is worth noting that {\cla} is not given any CL information whatsoever during both training and testing, and it treats each ground-truth subject and source equally without any CL-induced bias. This could explain the relatively flat yet high median \(P\)-values across all morphology classes. 
This suggests the CL-filtered sampling process described in Section \ref{sec:cl} is appropriate and does not introduce systematic bias correlated to consensus levels as far as training {\cla} is concerned.

\subsection{Model capacity and overfitting}
To investigate the impact of the large number of trainable parameters (over 136 million) on model overfitting given the relatively small training set (6,141 subjects), we conduct two experiments. In the first experiment, we reduce the number of model parameters from 136 million to 23 million, and in the second one, we reduce it further to 18 million. This is achieved by reducing the dimension of the two fully connected layers (Layer 24 and Layer 26 in Table \ref{tab:network_architecture}) from 4,096 to 256 and 64 respectively. We re-train these two `small-capacity' networks using the same training set (6,141 subjects), and test them against the same testing set (4,603 subjects). Their test accuracy --- mAPs of 82.9\% and 81.7\% --- is slightly poorer than {\cla} (mAP 83.6\%). This suggests that the {\cla} model is not in the overfitting zone~\citep{Goodfellow16}, in which higher model capacities correspond to higher test errors (thus lower test accuracy). We contend that the following factors mitigate overfitting in {\cla}.

First, the sole purpose of the two Dropout layers (Layer 25 and 27 in Table \ref{tab:network_architecture}) is to prevent {\cla} from overfitting during training, which is discussed in Section \ref{sec:rec_net}. Second, Section \ref{sec:loc_net} shows that, for each training subject, {\cla} dynamically generates thousands of RoI proposals and anchors to train the morphology classifier and the RoI regressor. This means the actual number of training examples going through the classification and regression loss functions (i.e. Eq \ref{eq:loss_anch_cls}, \ref{eq:loss_anch_reg}, \ref{eq:loss_roi_cls}, and \ref{eq:loss_roi_reg}) is on the order of millions rather than thousands. Moreover, transfer learning (discussed in Section \ref{sec:conv_net}) ensures that all parameters in the convolutional layers have been trained on the ImageNet dataset with millions of training images. This is particularly relevant to those `frozen' parameters in the low-level convolutional kernels.

For the above reasons, it is not essential to use other data augmentation techniques (such as image rotation) to enlarge the training set. More importantly, rotating an image around the source center, as is done in \citet{Aniyan17}, is not directly applicable to {\cla}. This is because it is {\cla}'s job to find sources on an image. The pre-processing step cannot possibly `reveal' a source $S$, and rotate/crop the image around $S$ since $S$ (and its location) is the very target {\cla} needs to predict. It is possible to blindly rotate the entire image/field around its own center regardless of the location of the sources. However, doing so may place some components of an extended source out of the field if we do not re-size the rotated image based on the rotation angle. Moreover, coordinates of the `new' corners of all boxes on the image need to be re-calculated and updated in the training set. Considering the above overheads, we will instead use {\cla}'s Spatial Transformer Network layer to support rotation invariant feature extraction as discussed in Section \ref{sec:implications} for our future work.

\subsection{Comparison with \citet{Aniyan17}}
The classifier in \citet{Aniyan17} produces a catalogue, which consists of 187 radio sources and their associated FR and BT morphology classifications (for both ground-truths and predictions) and spatial locations. A direct comparison between this catalogue and the {\cla} output is not feasible since the morphology categories used in \citet{Aniyan17} and {\cla} are different as described in Section \ref{sec:deep_learning}. However, if we consider all FRII sources having two radio components, it is possible to make an indirect comparison between the 57 FRII sources (out of the 187 sources) and all 2-component sources (i.e. 2C\_2P and 2C\_3P) predicted by {\cla}.

Out of the 4,603 subjects in the {\cla} testing set, the D3 method identifies 904 2-component sources, and the D4 method identifies 1,031 2-component sources. All the identified sources have $P$-values above 80\%. For each identified 2-component source, we calculate its center sky location from its bounding box coordinates predicted by {\cla}. This produces two location lists $L_3$ and $L_4$ for D3 and D4 respectively. We then perform spatial cross-matchings between $L_{3/4}$ and the 57 FRII sources predicted by the \citet{Aniyan17} classifier. When setting the maximum match radius to 20 arcsec, the cross-matching finds 1 match between the 57 FRII sources and $L_3$, and 1 match between the 57 FRII sources and $L_4$. Both matches are under a 3.5 arcsecond-radius, and both matches refer to the same pair: source \texttt{3C 251} in the \citet{Aniyan17} catalogue, and source \texttt{FIRST J110836.2+385854} in RGZ DR1. While the \citet{Aniyan17} classifier predicts it as an FRII source with a probability of 99.9\%, both D3 and D4 methods predicts its morphology as 2C\_3P with $P$-values of 95.7\% and 97.9\% respectively.

\section{Directions and recommendations for use of \cla}
\label{sec:empirical_eval}
We encourage interested astronomers to use \cla\ for their own research projects, because it can provide useful results even in its initial incarnation, and because experimentation and feedback on \cla\ will improve its performance.  Access to \cla's source code from the {\texttt{GitHub}} repository is described in Footnote~1.  Given the results to date, we recommend the use of either method D3 or D4. Therefore data would need to be provided in those forms, which can be obtained by following the descriptions on the {\texttt{GitHub}} repository. Also available in the repository are software modules that convert pairs of radio and infrared maps to these forms.

In Section~\ref{sec:assumption}, we describe how  \cla\ could be implemented in a simple automated manner for radio source classifications.  In Section \ref{sec:limitation_insights} we describe a variety of limitations in the current implementation, and in particular, how that would affect interpretation of the results.

\subsection{Classifying radio sources automatically with \cla}
\label{sec:assumption}

\subsubsection{How to use \cla?}
\label{sec:how_to_use}
For each input field, {\cla} detects and classifies the detected radio sources
 into the six RGZ morphology classes discussed in Section~3. Each classification
 generated will have a $P$-value which approximates the probability the 
identified source belongs to the identified morphology class. Therefore, 
 {\cla} may provide more than one morphology classification for each radio
source in the field.  An additional post-processing filtering algorithm is then recommended 
for deciding how to handle multiple classifications for a single radio source, as well as dealing with fields with more than one radio source.

The simplest filtering algorithm that a user can implement is to 
make two simple assumptions: 1) reject all classifications with $P$-values
below 0.8 unless the classification with the highest $P$-value is below 0.8; 
2) that there is only {\em{one}} radio source per field. While multiple
 sources exist within a test subject, our experience suggests that the source 
classification with the highest $P$-value is likely the correct classification 
as determined by {\cla}.  The assumption of one radio source per field is 
not unreasonable because 98.5\% of RGZ DR1 fields contain only one radio source (Wong et al., in preparation).
Further discussion on the impact of these assumptions can be found in  Section \ref{sec:remove_assumption}.

\subsubsection{Does this work?}
\label{sec:does_this_work}
The reliability analysis in Section \ref{sec:evaluation} does not include the filtering method described in Section \ref{sec:how_to_use}. From the perspective of an astronomer, the analysis in Section \ref{sec:evaluation} may not be sufficient because it is crucial for an astronomer to identify the correct classification from the multiple classifications produced by \cla. As such, we 
will describe, in this subsection, the accuracy and reliability of \cla\ in combination with the simple filtering method described in Section \ref{sec:how_to_use}.

To demonstrate that \cla\ (plus filtering) yields accurate and reliable classifications of 
resolved radio morphologies, we visually inspect an arbitrary sample of 500 test fields (from the entire testing set of 4603). We then apply the filtering method described in Section \ref{sec:evaluation} to this sample.
This arbitrary sample was selected via a simple Monte Carlo method that stops after a sample of 500 is reached. A plot that includes both RGZ DR1 classifications and {\cla} predictions is generated for each one of the 500 fields,  
which are then inspected and evaluated by a radio astronomer (OIW).  
367 of the 500 verification fields contain extended, non-compact radio sources, and 133 fields contain compact unresolved radio sources.

A mismatch between \cla\ and RGZ DR1 does not necessarily mean that one or the other is incorrect for two main reasons. Firstly, both \cla\ and RGZ classifications are limited by observational factors such as surface brightness sensitivity and resolution.  In addition, a mismatch in number of peaks can also be due to the limitation of the DR1 pipeline. Therefore a direct comparison between the classifications from \cla\ and those from DR1 is not a fair assessment of \cla's true performance. As such, we compare the results from \cla\ using the simple method described in Section \ref{sec:how_to_use} to RGZ DR1, and to a plausibility factor that is determined by an astronomer.  The main idea for the plausibility factor is to determine whether a classification from \cla\ can be deemed plausible by an expert astronomer given the radio and infrared maps presented, irrespective of the classification from the DR1 catalogue. For example, a field containing two unresolved radio source components with no infrared counterpart in between, or at the positions of the radio components, can be plausibly classified as either one 2C\_2P source or as two 1C\_1P sources.

We use a simple scoring method for quantifying the efficacy of \cla.  A score of one is awarded to each correct radio source classification. The total number of correct classifications is then divided by the total number of sources within the field. Hence, a field with multiple source classifications  will require a correct classification for each source to recover the total score of one for that field.  In this verification process, we ask two questions: 1) Does {\cla} reproduce the RGZ DR1 classification?; and 2) if {\cla} provides a classification \(C\) different from that of RGZ DR1, is \(C\) still plausible given the radio and infrared observations?  

Table~\ref{tab:visual_prop} lists the recovered verification scores for the 500 fields. Comparing the results from the D3 and D4 training methods to RGZ DR1, we find D4 to outperform D3 in a consistent manner.  While this is not surprising, it confirms that this scoring method works.  Taking into consideration the plausibility factor, our results show that {\cla} is likely to produce accurate source classifications at the optimal accuracy level above 93.1\% and 95.4\% using the training methods of D3 and D4.  Hence, we can expect reliable results from the current D4 version of \cla\ in combination with the simple post-processing filtering method described in Section \ref{sec:how_to_use}.


\begin{table}
\centering
\caption{Visual inspection results for the 500 verification fields for \cla's D3 and D4 training methods. We refer to the independent visual verification conducted by the radio astronomer which includes the plausibility factor, as `astronomer' in this table.}
\label{tab:visual_prop}
\begin{tabular}{lcc}
\hline
Compared to & D3 & D4\\
\hline \hline
RGZ DR1 & 447.0 & 465.2 \\
Astronomer & 465.5 & 477.2\\
\hline
\end{tabular}
\end{table}

\subsection{Limitations and insights}
\label{sec:limitation_insights}
While Section \ref{sec:assumption} shows that \cla\ is a relatively accurate and reliable prototype classifier, we caution the reader and users of \cla\ that the current version does include
a number of limitations that we discuss in more details in this subsection.
Previously in Section \ref{sec:does_this_work}, we noted that 
a mismatch between the two does not necessarily mean that either \cla\ or RGZ is incorrect.
In this subsection, we explore and describe the limitations and lessons learnt from the
implementation of \cla, from the perspective of an astronomer.
 
There are several reasons why a mismatch between the two methods may still 
result in a plausible source classification.  For many complex radio sources, 
further follow-up observations may be required to ascertain the precise 
source component associations and host galaxy.  Furthermore, the determination
of the number of peaks is an approximation by the DR1 pipeline that is based upon
the contour levels.  
Hence we discuss in Section \ref{sec:remove_assumption}, 
 \cla's reliability from an astronomer's perspective based on the 
often-used method of visual inspection.

\subsubsection{Source angular size}
\begin{figure*}
	\includegraphics[scale=0.74]{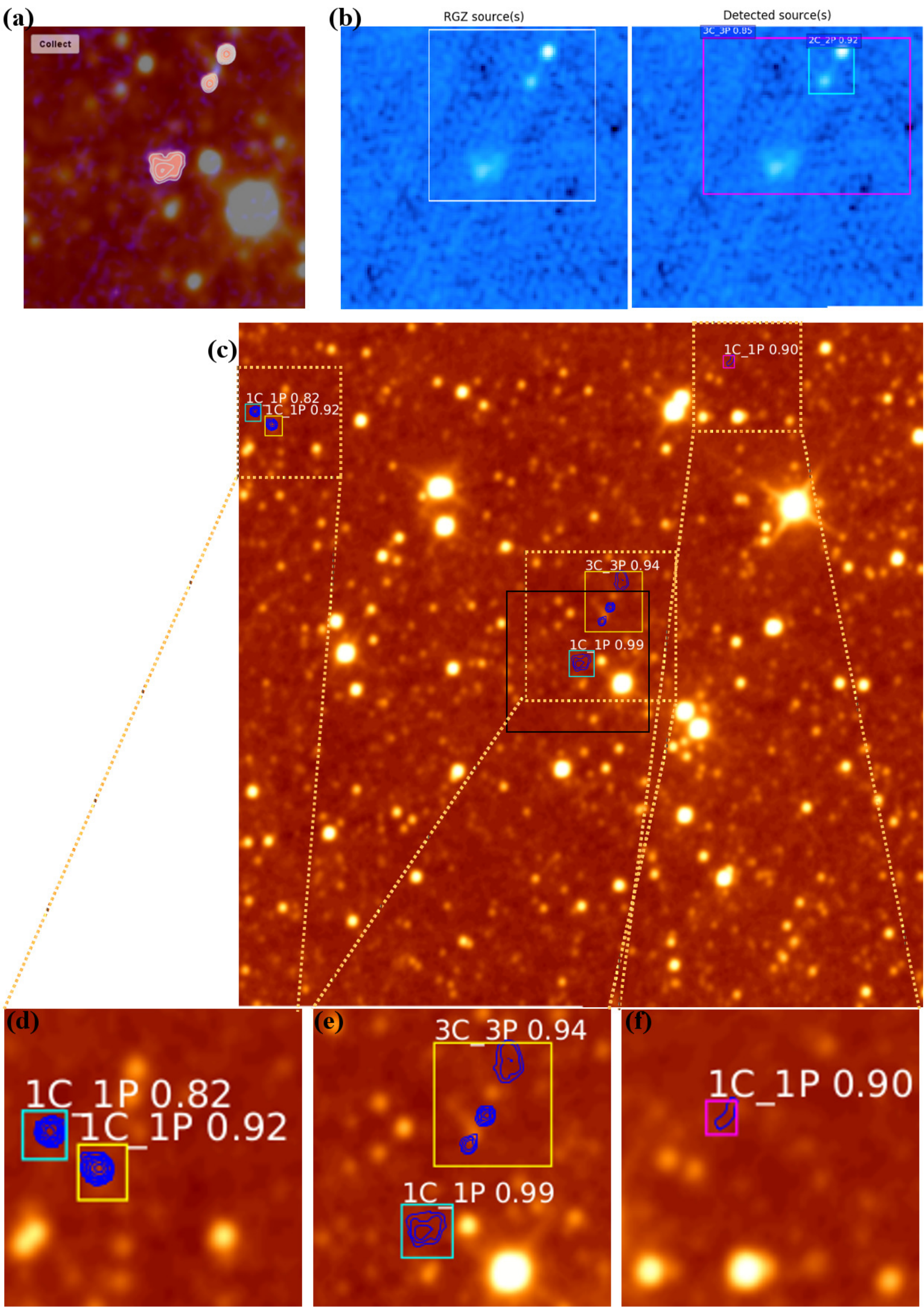}
    \caption{An example of the limitation inherent to the 3-arcmin FoV whereby radio sources are misclassified by both RGZ DR1 and \cla\ due to the missing radio component that lies beyond the Northernmost-edge of the field. North and East are to the top and left of the page, respectively.  Panel (a) shows the 3-arcmin by 3-arcmin RGZ subject presented to the participants for RGZ J080837.0+170804. Panel (b) is a pair of verification maps showing the DR1 classification (left) and \cla's classification (right). Panel (c) shows the expanded 15-arcmin by 15-arcmin FoV for  RGZ J080837.0+170804 where the original RGZ image size is marked by the black box.  For added visual clarity, zoomed-in maps of the three radio sources found by \cla\ within Panel (c) are shown in panels (d), (e) and (f). }
    \label{fig:egfovlimit}
\end{figure*}

Similar to the Radio Galaxy Zoo project, \cla\ will not be able to provide accurate classifications for radio sources which extend beyond the 3-arcmin FoV. RGZ DR1 found the median angular size of multicomponent radio sources to be  43.1 arcseconds and that 95.2\% of the DR1 multicomponent sources have an angular size that is smaller than 97 arcseconds (Wong et al., in preparation).  However, there is a small fraction of sources which may be limited by the current FoV size.  Figure~\ref{fig:egfovlimit}a illustrates one example field within the verification set of 500 that encounters the limitations of the 3-arcmin FoV, whereby the field presented in RGZ only encapsulates three of the four radio components. The northern-most radio component lies beyond the top-edge of the field.  Consequently, both the classifications from RGZ DR1 and \cla\ are incorrect (Figure~\ref{fig:egfovlimit}b). Enlarging the field by five times to a 15~arcmin by 15~arcmin field (Figure~\ref{fig:egfovlimit}c), we reveal that the central radio source has a double-double morphology (4C-4P), for which \cla\ was not trained to identify. When running directly on this larger field, {\cla} ends up breaking this double-double source into two smaller sources --- 3C-3P and 1C-1P (Figure~\ref{fig:egfovlimit}e). On the other hand, the host galaxy captured inside the 3C-3P bounding box is still correct.

Of the 500 verification fields, we find two classifications in which \cla\ estimated a significantly larger angular source size (by a factor of a few) relative to that reported by RGZ DR1. Two most likely reasons exist for such an estimation: either \cla\ is confused by the synthesis imaging artefacts that remain in some fields, or that \cla\ is capable of detecting low level diffuse emission. We will investigate these specific aspect of \cla\ in future studies as it is beyond the scope of this proof-of-concept paper to provide an in-depth investigation into this specific area.

\subsubsection{Assumption of one source per field} 
\label{sec:remove_assumption}
Of the two assumptions recommended for the simple filtering method in Section \ref{sec:how_to_use}, the second assumption of one source per field, may not be necessary for some studies to obtain
individual classifications.  Also, this assumption of one source per field may be invalidated for two main reasons.  Firstly,  multicomponent radio sources with large angular sizes can result in multiple plausible classifications as discussed in the previous subsection.  Secondly, the classifications of multiple radio sources in the 8\% of verification fields are not distinguishable from multiple classifications
of a single multicomponent source. Hence this subsection investigates the reliability
of \cla\ when we remove the single-source assumption.

\begin{table}
\centering
\caption{The fraction of verification fields within three divisions of completeness ratios.}
\label{tab:cratio_prop}
\begin{tabular}{lcc}
\hline
$N_{\rm{TRUE}}/N_{\rm{CLASSIFIED}}$ & $D3$ & $D4$\\
\hline \hline
$=1$ & 0.834 & 0.864 \\
$<1$ & 0.154 & 0.122\\
$>1$ & 0.012 & 0.014\\
\hline
\end{tabular}
\end{table}

\begin{figure}
    \begin{tabular}{c}
    	\includegraphics[width=\columnwidth]{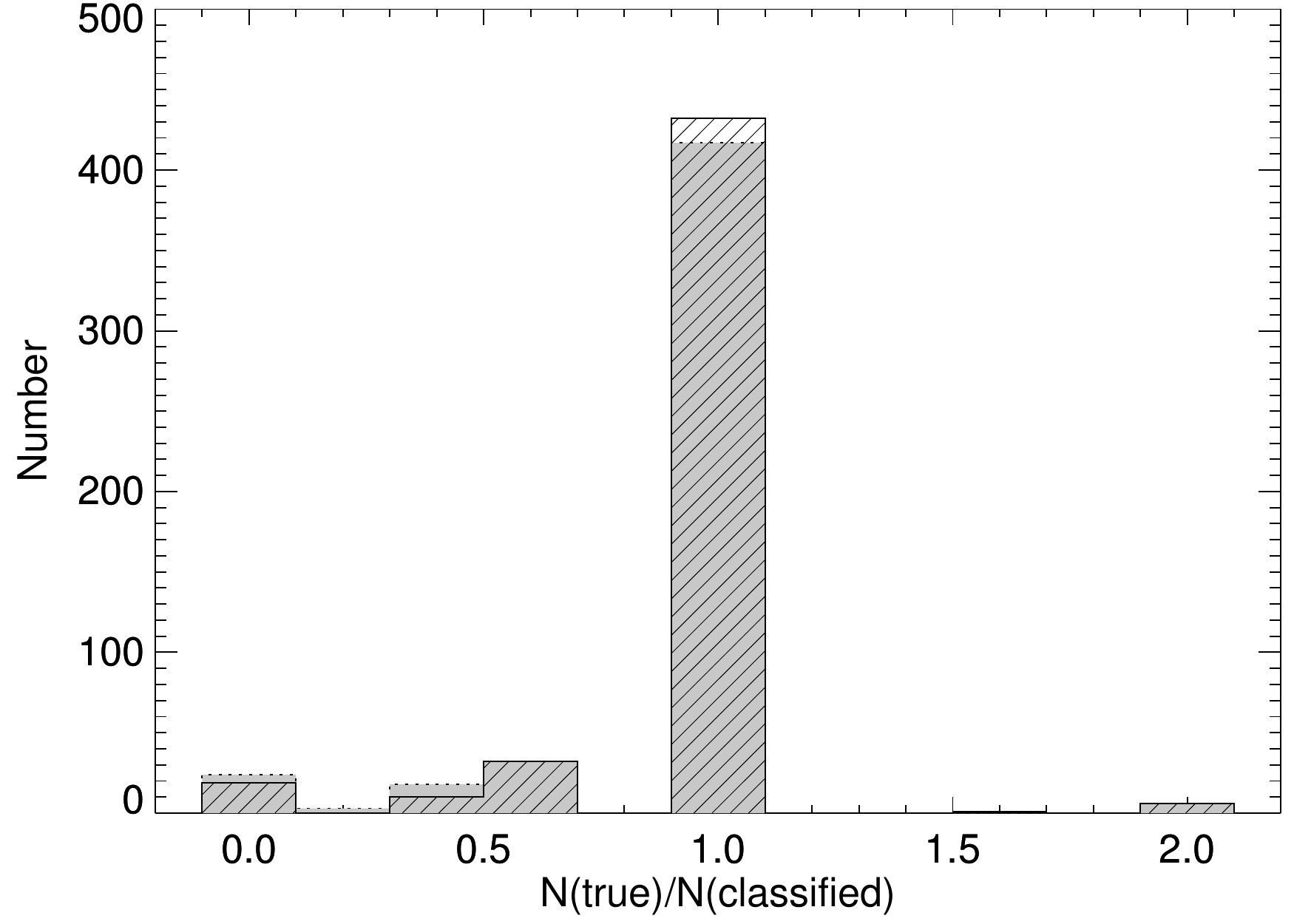}\\
    	\includegraphics[width=\columnwidth]{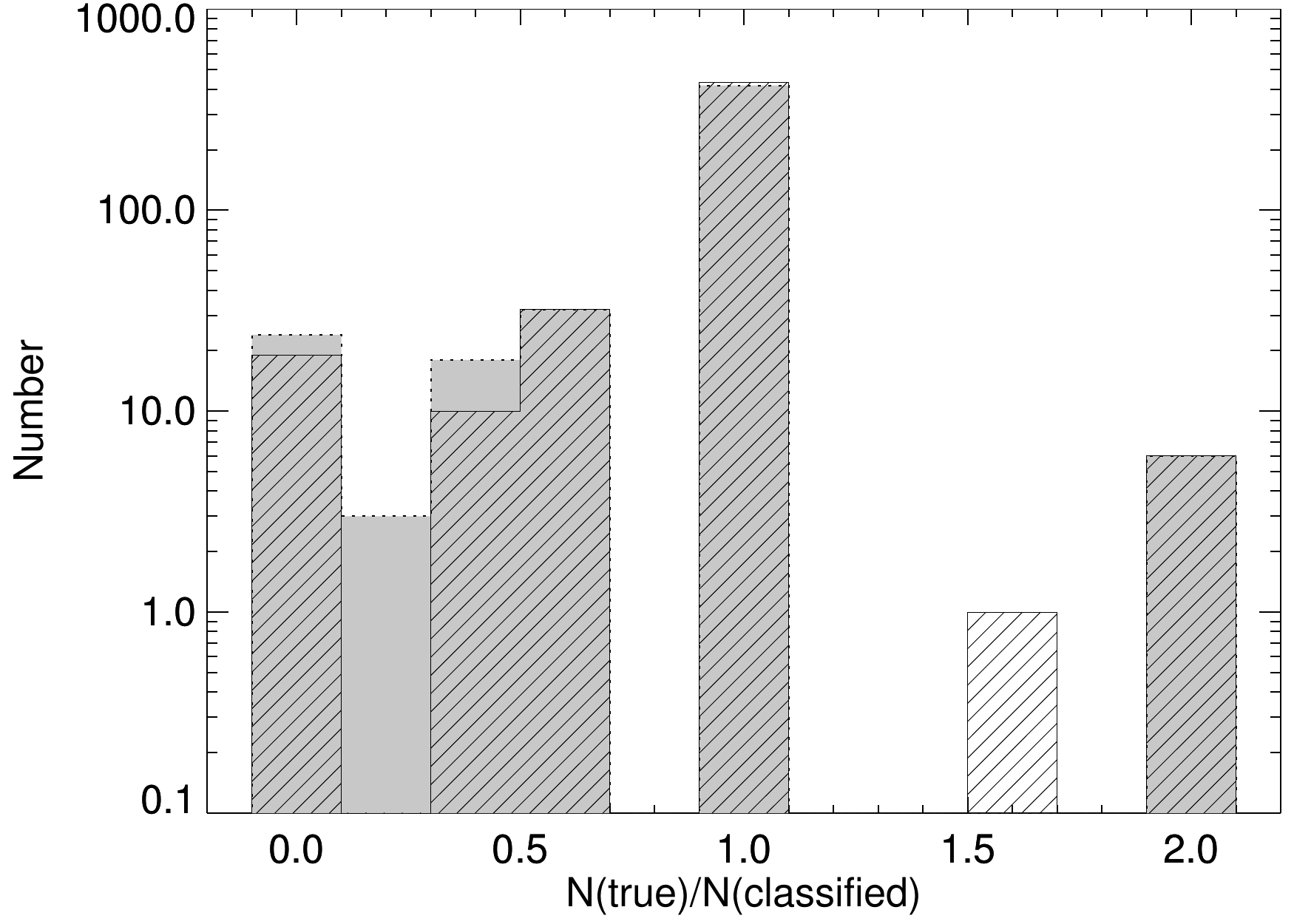}\\
    \end{tabular}
    \caption{Fraction of total number ($N_{\rm{true}}$) to classified number ($N_{\rm{classified}}$) of radio sources in the verification sample of 500.  The top figure shows a linear y-axis and the bottom figure shows a logarithmic y-axis for the same distribution. Verification results from the D3 and D4 methods are represented by the grey shaded and striped distributions, respectively. }
    \label{nfracdistrib}
\end{figure}

To this end, we examine the classification degeneracies that become inherent (when we do not assume one source per field) using a completeness ratio.  We define a \textit{completeness ratio} to be the ratio of the total number of radio sources to the total number of correct classifications per subject ($N_{\rm{TRUE}}/N_{\rm{CLASSIFIED}}$).  A ratio of 1.0 indicates that every source within a field has been correctly classified. Ratios greater than 1.0 suggest that an individual field contains more DR1 sources than classified. Likewise, a ratio below 1.0 indicates that \cla\ found more than one classification per source within a field. Figure~\ref{nfracdistrib}
 presents the distribution of completeness ratios for the 500 verification fields using the D3 (grey-shaded) and D4 (striped) training methods. As shown in Table~\ref{tab:cratio_prop}, we find that a ratio of 1.0 is obtained for 83.4\% and 
 and 86.4\% (using the D3 and D4 methods, respectively) of the verification sample. This result is consistent with the precision of the classifications quantified in Section \ref{sec:evaluation}. However, since out of the 500 verification fields, only 36 are in fact multi-source subjects, and the majority (464) of them still contain only one source per FoV. Therefore Table~\ref{tab:cratio_prop} may not generalize well beyond this particular 500-subject sample to reflect the effect of `removing the single-source assumption' on {\cla}'s reliability. To address this issue, we create another special subset $S$ (\(|S|=36\)) from the 500 verification fields, in which each subject has at least two sources. We re-calculate the completeness ratios against $S$, and report the updated result in Table \ref{tab:cratio_36_prop}. Compared to Table~\ref{tab:cratio_prop}, Table \ref{tab:cratio_36_prop} essentially examines classification degeneracies under the `worst-case' scenario where every subject has multiple sources. We find that the updated results of 55.6\% and 66.8\% (for a completeness ratio of 1.0) are largely consistent with Table \ref{tab:eval_multisource}.



\begin{table}
\centering
\caption{The fraction of verification fields on a set of 36 subjects, each of which has at least two sources in its FoV}
\label{tab:cratio_36_prop}
\begin{tabular}{lcc}
\hline
$N_{\rm{TRUE}}/N_{\rm{CLASSIFIED}}$ & $D3$ & $D4$\\
\hline \hline
$=1$ & 0.556 & 0.668 \\
$<1$ & 0.278 & 0.166\\
$>1$ & 0.166 & 0.166\\
\hline
\end{tabular}
\end{table}

\section{Discussion}
\label{sec:implications}

This paper builds upon earlier machine learning explorations that use RGZ classifications as a training set \citep{lukic18,alger18}.  Following \citet{alger18} who found that  compact radio source classifications do not benefit significantly from using advanced machine learning convolution methods, we specifically train and test \cla\ on a large sample of extended non-compact radio sources. Our work demonstrates the feasibility of applying modern deep learning methods, which originate from generic object detection and computer vision, for cross-matching complex radio sources of multiple components with infrared catalogues. The promising results of this study have implications for further development of fully-automated cross-wavelength source identification, matching, and morphology classifications for pre-SKA surveys.
The data fusion methods and their performance evaluations described in this paper provide a good starting point to train machines to appropriately incorporate and integrate numerous deep multi-wavelength catalogues and other information (e.g. redshifts) for radio source identification and morphology classification.

We adapt the Spatial Transformer Network (STN) for cropping out RoI proposals from feature maps in order to obtain a differentiable loss function for end-to-end training. The adaptation takes place in the affine transformation matrix (Eq. \ref{eq:affine_t}) where we fix the rotation angle to zero degree (thus no rotation). 
By running a fully-fledged STN that allows rotation angles to be learned from the feature map,  
{\cla} could perform source-dependent, rotation-invariant morphology classification within a single end-to-end pipeline. This approach will differ from random rotations of the entire image for feature augmentation ~\citep{dieleman2015rotation} because it is trained to rotate each potential source by a distinct angle for optimal morphology classification. The assumption that all sources within a subject rotate simultaneously by a pre-defined angle does not always hold. 

Despite the great difference between common images and RGZ subjects, we demonstrate that the CNN weights thoroughly trained on the comprehensive, well-labelled ImageNet provide far better initial conditions than random weight initialization with respect to training efficiency and evaluation metrics. However, as shown in Figure \ref{fig:learning_curve}, appropriate control of these pre-loaded weights is equally important in order to achieve a desirable level of efficiency and precision. The fact that freezing low-level weights leads to a much smaller training error suggests that high-level feature extractions (such as shapes, texture, structure, etc.) should be prioritized after transfer learning. 

On the other hand, low-level feature learning should be carried out at a much slower pace to avoid over-fitting. This implies that we may need different learning rates for different parts of the neural network. In this example, freezing weights is equivalent to reducing the learning rate to 0, which simply gives up opportunities to learn any low-level features unique to the RGZ dataset. Therefore, a more fine-grained learning rate distribution applied across the network can take advantage of the benefits from transfer learning.

\section{Conclusions}
\label{sec:conclusions}
Cross-identification of radio source components 
is currently done through visual inspection by expert astronomers or citizen scientists. However, such a labour-intensive method is not scalable even for the pre-SKA (Square Kilometre Array) radio surveys such as EMU --- Evolutionary Map of the Universe~\citep{norris11}. In this paper we describe a machine learning-based method for automated localization and identification of multi-component, multi-peak radio sources with associated morphological information. Drawing on the latest models developed in object detection and deep learning, our method has achieved efficient identification of radio galaxies on unseen RGZ datasets with an mean Average Precision (mAP) of 83.6\% and an empirical plausibility accuracy of above 90\%. {\cla} is able to distinguish between six of the most common distinct classes of radio source morphologies. These six classes of morphologies are defined in terms of the number of components and peaks that describe source associations and identifications produced by the Radio Galaxy Zoo Data Release 1 (Wong et al., in preparation). \cla\ also works reasonably well on fields much larger than those provided in the training set.

For future work, we will focus on improving {\cla}'s capability of separating unrelated multiple sources from multiple components of single sources. To begin with, we will incorporate more multi-source subjects into the testing set as suggested by Table \ref{tab:eval_multisource}. Targeted plausibility analysis of the confusion between 2C and 3C classifications and reliable statistical measures will help us develop robust feature augmentation schemes needed to address this key problem.

\section*{Acknowledgements}
We acknowledge the contribution of more than 12,000 volunteers in the Radio Galaxy Zoo project. 
The International Centre for Radio Astronomy Research (ICRAR) is a joint venture between Curtin University and The University of Western Australia with support and funding from the State Government of Western Australia. This work was supported by computing resources provided by the Pawsey Supercomputing Centre (with funding from the Australian Government and the Government of Western Australia) and CSIRO. Partial support for LR's and AG's work at the University of Minnesota comes from grant AST-1714205 from the U.S. National Science Foundation. HA benefited from grant DAIP \#66/2018 of Universidad de Guanajuato.




\bibliographystyle{mnras}
\bibliography{rgzcitations} 

\label{lastpage}
\end{document}